\documentclass[sigconf]{acmart}

\usepackage{graphicx}
\usepackage{subcaption}
\usepackage{pifont}
\newcommand{\cmark}{\ding{51}}%
\newcommand{\xmark}{\ding{55}}%
\usepackage{multirow}
\usepackage{pdfpages}
\usepackage{longtable}

\AtBeginDocument{%
  \providecommand\BibTeX{{%
    \normalfont B\kern-0.5em{\scshape i\kern-0.25em b}\kern-0.8em\TeX}}}

\copyrightyear{2021}
\acmYear{2021}
\setcopyright{acmcopyright}\acmConference[CHI '21]{CHI Conference on Human Factors in Computing Systems}{May 8--13, 2021}{Yokohama, Japan}
\acmBooktitle{CHI Conference on Human Factors in Computing Systems (CHI '21), May 8--13, 2021, Yokohama, Japan}
\acmPrice{15.00}
\acmDOI{10.1145/3411764.3445683}
\acmISBN{978-1-4503-8096-6/21/05}

\begin{document}

\title{Effective Interfaces for Student-Driven Revision Sessions for Argumentative Writing}


\author{Tazin Afrin}
\affiliation{%
  \institution{University of Pittsburgh}
  \streetaddress{270 S Bouquet St.}
  \city{Pittsburgh}
  \country{USA}}
\email{tazinafrin@cs.pitt.edu}

\author{Omid Kashefi}
\affiliation{%
  \institution{University of Pittsburgh}
  \streetaddress{270 S Bouquet St.}
  \city{Pittsburgh}
  \country{USA}}
\email{kashefi@cs.pitt.edu}

\author{Christopher Olshefski}
\affiliation{%
  \institution{University of Pittsburgh}
  \streetaddress{270 S Bouquet St.}
  \city{Pittsburgh}
  \country{USA}}
\email{cao48@pitt.edu}

\author{Diane Litman}
\affiliation{%
  \institution{University of Pittsburgh}
  \streetaddress{270 S Bouquet St.}
  \city{Pittsburgh}
  \country{USA}}
\email{litman@cs.pitt.edu}

\author{Rebecca Hwa}
\affiliation{%
  \institution{University of Pittsburgh}
  \streetaddress{270 S Bouquet St.}
  \city{Pittsburgh}
  \country{USA}}
\email{hwa@cs.pitt.edu}

\author{Amanda Godley}
\affiliation{%
  \institution{University of Pittsburgh}
  \streetaddress{270 S Bouquet St.}
  \city{Pittsburgh}
  \country{USA}}
\email{agodley@pitt.edu}

\renewcommand{\shortauthors}{Afrin et al.}

\begin{abstract}
We present the design and evaluation of a web-based intelligent writing assistant that helps students recognize their revisions of argumentative essays. To understand how our revision assistant can best support students, we have implemented four versions of our system with differences in the unit span (sentence versus sub-sentence) of revision analysis and the level of feedback provided (none, binary, or detailed revision purpose categorization). 
We first discuss the design decisions behind relevant components of the system, then analyze the efficacy of the different versions through a Wizard of Oz study with university students.
Our results show that while a simple interface with no revision feedback is easier to use,  
an interface that provides a detailed categorization of sentence-level revisions is the most helpful based on user survey data, as well as the most effective based on improvement in writing outcomes.


\end{abstract}

\ccsdesc[500]{Human-centered computing~Interactive systems and tools}
\ccsdesc[300]{Human-centered computing~Graphical user interfaces}
\ccsdesc[300]{Human-centered computing~Web-based interaction}
\ccsdesc[300]{Human-centered computing~Natural language interfaces}
\ccsdesc[100]{Human-centered computing~Empirical studies in interaction design}
\ccsdesc[500]{Human-centered computing~Empirical studies in HCI}
\ccsdesc[500]{Applied computing~Education}
\ccsdesc[300]{Computing methodologies~Artificial intelligence}

\keywords{Academic writing; revision; argumentative writing; intelligent interface; wizard of oz}


\maketitle

\section{Introduction}
Argumentative writing has long been considered a key component in academic and professional success. Educational research has established that not only does argumentative writing produce positive learning gains among students, but it also contributes to more complex critical thinking skills~\cite{fitzgerald2000reading,newell2014HighSE}. 
However, many students lack the skill of developing an argumentative essay without any writing instruction. 
Typically, instruction of argumentative writing involves both the composition of multiple drafts of writing and revising those drafts based on formative feedback from others (e.g. teachers, peers). Although most educators and writing instructors agree on the importance of formative feedback, teachers have observed that it can be especially time-consuming, and are thus challenged to consider the balance between efficacy and efficiency ~\cite{paulus1999}. Research on peer feedback suggests that students often do not benefit from peer responses unless peer reviewers have been explicitly instructed how to do it~\cite{loretto2016secondary}. 

As a solution, scholars of Natural Language Processing (NLP) have worked toward developing automated writing assistant tools in order to provide instant and constructive feedback to student writers. 
Many of these tools, however, provide product-focused feedback for one draft at a time (e.g. essay scoring ~\cite{attali2006b}, error correction ~\cite{grammarly}, argument mining~\cite{chernodub2019targer}), as opposed to process-focused feedback, which could provide writers with information not only on the quality of a single draft of writing, but also on the evaluation of their revision patterns from previous to the current draft of an essay. The idea behind ArgRewrite\footnote{http://argrewrite.cs.pitt.edu/}, 
the tool described in this paper, is that improving as a writer involves not only producing increasingly higher quality writing, but it also involves improving on the way one engages in the revision process. 
The ArgRewrite is designed to help students iteratively revise and update their essays. While previous work shows that feedback on textual revisions encourages students to further revise their essays ~\cite{zhang2015l, zhang2017hh}, in this study we want to understand the level of revision categorization (e.g., binary versus detailed) and unit of analysis (sentence or sub-sentential) that is most effective in helping students improve their essay. We hypothesize that a more detailed categorization of a student's revision would be more useful. With that in mind, we design four web-based interface conditions of the ArgRewrite revision assistant tool -- ranging from control with no revision categorization to sentence-level and sub-sentential revision categorization. 



This article presents data from a lab-based experiment in which users were provided with one of four different versions of the web-based ArgRewrite tool, each of which differs in unit span of revision analysis and levels of detail in the revision purpose categorization. Condition A is our control interface which provides no feedback at all. Condition B provides binary revision categorization for sentence-level revisions, 
condition C provides detailed revision categorization for nine different types of sentence-level revisions, and finally condition D used the same revision categorization as C, but provided categorization for sub-sentential revisions. First, we describe the interface components and design decisions for each condition of the ArgRewrite. To understand the usefulness of each condition, we then look at student perception of the system by analyzing the user survey about the interface. Our analysis shows that although our conditions with feedback are not always easy to use compared to the simple control condition, students find the revision categorization helpful to understand their revision effort and weakness. Especially, condition C with detailed sentence-level revision categorization showed to be most useful. Detailed revision categorization also encouraged students to make more revision, qualitatively and quantitatively. 
We also tested the effectiveness of the system in helping students to further improve their essay score. Again, detailed sentence-level categorization showed to be more useful in helping students boost the essay score. Our research contributions are four fold: 

\begin{itemize}
    \item We developed four conditions of an argumentative revision assistant tool that supports different levels of revision feedback (e.g., binary versus detailed purpose categorization; sentence versus sub-sentential revision unit) and conducted a lab-based study, where students used the tool to revise their essays.
    \item Using statistical analyses, we compare the usability of the conditions of the tool to understand the revision feedback most helpful from a user perspective.
    
    \item Using statistical analyses, we compare the essay score gain to understand what is the best revision feedback to help improve the essay.
    
    \item We categorize the revisions students made and perform a comparative analysis to understand the revision behavior by students using different conditions.
\end{itemize}


\section{Related Work}


Many of the NLP-based writing assistant tools that were developed over the last few years provide feedback on one writing product at a time, or focus on high-level semantic changes. 
For example, Grammarly~\cite{grammarly} provides feedback on grammar mistakes and fluency,  ETS-writing-mentor~\cite{ets-writing-mentor} provides feedback to reflect on higher-level essay properties such as coherence, convincingness, etc. Other writing assistant tools such as EliReview~\cite{elireview}, Turnitin~\cite{turnitin} are designed for peer feedback, plagiarism detection, etc., rather than focusing on writing analysis and feedback. In contrast to those existing tools, we compare two drafts using the ArgRewrite revision assistant tool. 
While a previous version of ArgRewrite~\cite{zhang2016hl} provided feedback based on detailed revision categorization ~\cite{zhang2015l,zhang2017hh} at the sentence-level and was evaluated via a user survey, the current study develops two additional ArgRewrite interfaces (based on binary revision categorization and sub-sentential revision units) and  evaluates all interfaces using both user survey and writing improvement analysis. 

In terms of revision analysis, work on Wikipedia is the most related to the study of academic writing. Prior works on Wikipedia revision categorization focus on both coarse-level~\cite{bronner2012m} and fine-grained~\cite{johnes2008wikirevision, daxenberger2012g, yang2017hkh} revisions. However, because some fine-grained Wikipedia categories (e.g., vandalism) are specific to wiki scenarios, writing studies instead use fine-grained revision categories more suitable for student argumentative writing~\cite{toulmin_2003,zhang2015l}.  In both cases (Wikipedia or educational), previous studies have focused on investigating the reliability of manually annotating and automatically classifying coarse-level and detailed revision categories, as well as on demonstrating correlations between category frequency and outcome measures. In contrast, our study manipulates whether ArgRewrite provides feedback using coarse-level  (surface versus content) or detailed  (e.g., claim, evidence, etc.) revision categorizations of textual changes. 

Previous studies on writing revision research vary as to whether they use the word-level ~\cite{bronner2012m,daxenberger2012g} or the sentence-level as the revision span~\cite{zhang2016hl}. Sentences represent a natural boundary of text and automatic revision extraction at the sentence-level has  been shown to be reasonably accurate ~\cite{zhang2014alignment}. 
However, sentence-level revision categories may not always be appropriate. For example, a sentence revision may contain a few fluency changes at the beginning, with substantial information added at the end. In that case, that sentence contains both surface and content revisions. With that in mind, in addition to the sentence-level revisions that were the focus of the original ArgRewrite~\cite{zhang2016hl}, the current study also explores sub-sentential revisions with detailed revision categorization. 



The writer's previous revision effort is often studied in collaborative writing to visualize revisions from multiple authors. For example, DocuViz ~\cite{wang2015docuviz} tracks the number of revisions in google docs and shows the pattern of revising and developing a collaborative document by multiple authors. Unlike collaborative writing, our work focus on multiple revisions by a single author. 
Another research work that studies visualizing multiple revision patterns by a single student also focuses on the amount of revision through an automated revision graph~\cite{shibani2018kb,shibani2020RevisionGraph}.
Although our ArgRewrite tool does show the number of revisions for each revision category, we do not categorize the revisions based on the frequency. Instead, the revision categories reflect the purpose~\cite{zhang2015l} of that revision. In our tool, the revision are highlighted in both drafts of the essay.




In argument mining, the main goal is to find argument structures and their relations from text. It also focuses on a single text. However, few tools are available for argument mining. One recent work experiments with a text editor to support the student argumentation skills~\cite{wambsganss2020CHIargskills}. The tool provides feedback on the argumentation quality of a given text. Students using the tool wrote a more convincing argument than students in the control/baseline condition. A tool called ArguLens helps find issues in issue tracking systems using automatic argument mining~\cite{wang2020CHIargulens}. Another recent tool for argument mining is called TARGER~\cite{chernodub2019targer}, which also visualizes argumentative phrases in a text of a single draft. Unlike these argument mining tools, our ArgRewrite focuses on argumentative revision ~\cite{zhang2015l} and compares two drafts of student essays.




Works on formative feedback usually focus on embedded writing instructions for students to further improve the article ~\cite{wingate2010impact,shute2008focus,knight2020acawriter}. 
While we provide revision analysis and show it with corresponding highlight colors on our web-based tool, this is not a study about providing formative feedback on student essays, or the quality of feedback. Rather, our study focuses on helping students to understand their previous revision effort, or how they addressed the feedback received on the previous draft of an essay. 
Monitoring one's own progress towards a goal is a cognitively complex task called self-regulation~\cite{ZimmermanBandura1994,ZimmermanKitsantas2002}. Previous studies have shown that self-regulation has a positive impact on students' writing development~\cite{macarthur2015self,ZimmermanBandura1994}. In our study, self-regulation occurs both during the reflection of previous revision efforts and during the actual revision process. 
Our ArgRewite tool does not suggest any future revision automatically. Instead, it presents its analysis (but not quality evaluation) of previous revisions so that students can make informed decisions when they further revise the essay.

\section{ArgRewrite System Overview}
\begin{figure}[!h]
  \centering
    \includegraphics[width=1.0\linewidth]{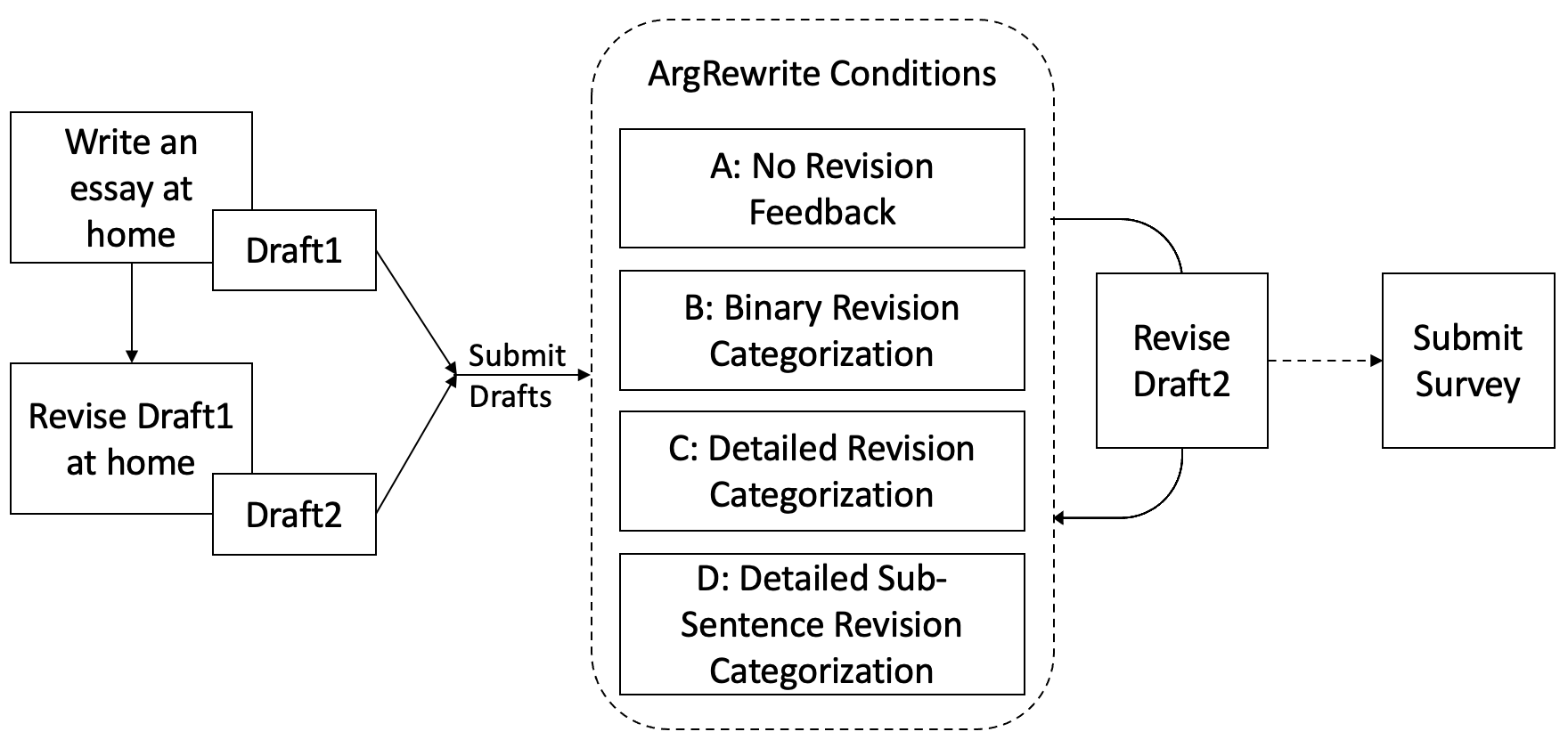}
  \caption{ArgRewrite essay revision process}
  \label{fig: process}
  \Description{ArgRewrite essay revision process for experimental study.}
\end{figure}

\begin{table}
  \caption{ArgRewrite interface conditions}
  \label{tab: cond comparison}
  \begin{tabular}{lcccc}
    \toprule
    ArgRewrite Conditions: &	A	&	B	&	C	&	D	\\ 
    \midrule
	Sentence-level Revision	&	\xmark	&	\cmark	&	\cmark	&	\xmark	\\ 
    Sub-sentence-level Revision	&	\xmark	&	\xmark	&	\xmark	&	\cmark	\\ 
    Binary Revision Categorization	&	\xmark	&	\cmark	&	\xmark	&	\xmark	\\ 
    Detailed Revision Categorization	&	\xmark	&	\xmark	&	\cmark	&	\cmark	\\ 
    \midrule
    Number of participants & 20 & 22 & 22 & 22 \\
    \bottomrule
  \end{tabular}
\end{table}


Figure~\ref{fig: process} shows the essay revision process using the ArgRewrite tool. Experimental participants were recruited through flyers targeting undergraduate and graduate-level students who were either native English speakers or non-native speakers with a certain level of English proficiency (TOEFL score > 100). In our experiment, there are two rounds of essay revision, Draft1 to Draft2, and Draft2 to Draft3. 
Participants wrote their first draft (Draft1) of an essay at home based on a given prompt\footnote{The prompt is provided in \ref{a sec: prompt and rubric}}. After a few days of finishing Draft1, each participant received expert feedback\footnote{The experts were a professor and a graduate student in the School of Education, and a trained undergraduate student. An example of expert feedback is provided in \ref{a sec: expert feedback}. The rubric that guided the feedback for Draft1 parallels exactly the scoring rubric (\ref{a sec: rubric}) other than the addition of the pronoun “you.”} 
on their essay argument quality and overall writing structure. Based on the feedback, they revised their Draft1 and produced Draft2. 
After finishing Draft2, participants were randomly assigned to use different conditions of the ArgRewrite in a lab environment. They did not receive any feedback on their Draft2. Instead, they are shown the ArgRewrite interface on a computer highlighting their previous revision from Draft1 to Draft2. Participants were asked to use the tool to revise their Draft2 and create a final and generally improved version of the essay, Draft3.

Although our tool supports full automation of revision categorization, we relied on Wizard-of-Oz prototyping ~\cite{browne2019chiWOZ} for this particular experiment. In Wizard-of-Oz prototyping, a human manually handles the automation, but the student cannot tell the difference from the web-interface they see. We did so to eliminate the confounding factors of NLP automation errors when we compare different conditions.
The background server of ArgRewrite uses NLP to automatically segment the essays into sentences and align the two drafts at the sentence-level~\cite{zhang2016hl}. Modified, added, or deleted sentences were then extracted as \textit{revisions}. 
The ArgRewrite server automatically extracts those revisions and classifies them into different revision purpose categories.
In our Wizard of Oz experimental setting, a human then fixes the server errors for alignment and classification before the participants start the second round of revision in the lab. In the lab-based experiment, participants first read a short tutorial on using the ArgRewrite tool.
Then they were asked to go through their previous revision effort. In conditions B, C, and D, they also submitted confirmation if they agree or disagree with the revision categories for each of the revised sentences the tool is showing them. They did so before and after completing the final revision. 
Finally, after the participants finished revising the essay, they were asked to answer survey questions about the interface. 

Table~\ref{tab: cond comparison} shows the main differences among the ArgRewrite conditions and the number of participants for each condition. 86 participants were assigned randomly for each condition. Out of 86 participants, 69 were native English speakers, and 17 non-native speakers. The number of non-native speakers in conditions A,B,C,D are 3,4,5,5 respectively. A separate study on participants' native speaking skills showed that non-native speakers made significantly more revisions than native speakers in the first round of revision but not in the second round. Although non-native speakers' scores were lower than native speakers on all drafts and in all conditions, there were no significant differences in non-native vs native speakers revisions or scores across conditions.



\section{Web-based Interface}
Drawing on research on learning analytics~\cite{liu2013Tracer-tool,verbert2013-learning-analytics}, ArgRewrite is designed to facilitate personal learning. 
According to Verbert et al.~\cite{verbert2013-learning-analytics}, learning analytics systems provide visualizations and overviews in order to make the users aware of relevant and important information. 
Each ArgRewrite condition has two parts - the overview interface and the rewrite interface.  
The overview interface gives a summary of students' revisions between the two submitted drafts, while the rewrite interface is where students revise their current draft.
Following the previous study~\cite{zhang2015l}, in the case of ArgRewrite, the overview interface was designed to bring users' awareness of the purpose of their latest revisions. Then on the rewrite interface, they were asked to go through each revision label to determine whether or not the system identified their revision purposes correctly. Finally, users were allowed to further revise their essay to improve the overall quality. 

\subsection{Overview Interface}
\begin{figure}[h]
  \centering
    \fbox{\includegraphics[width=0.95\linewidth]{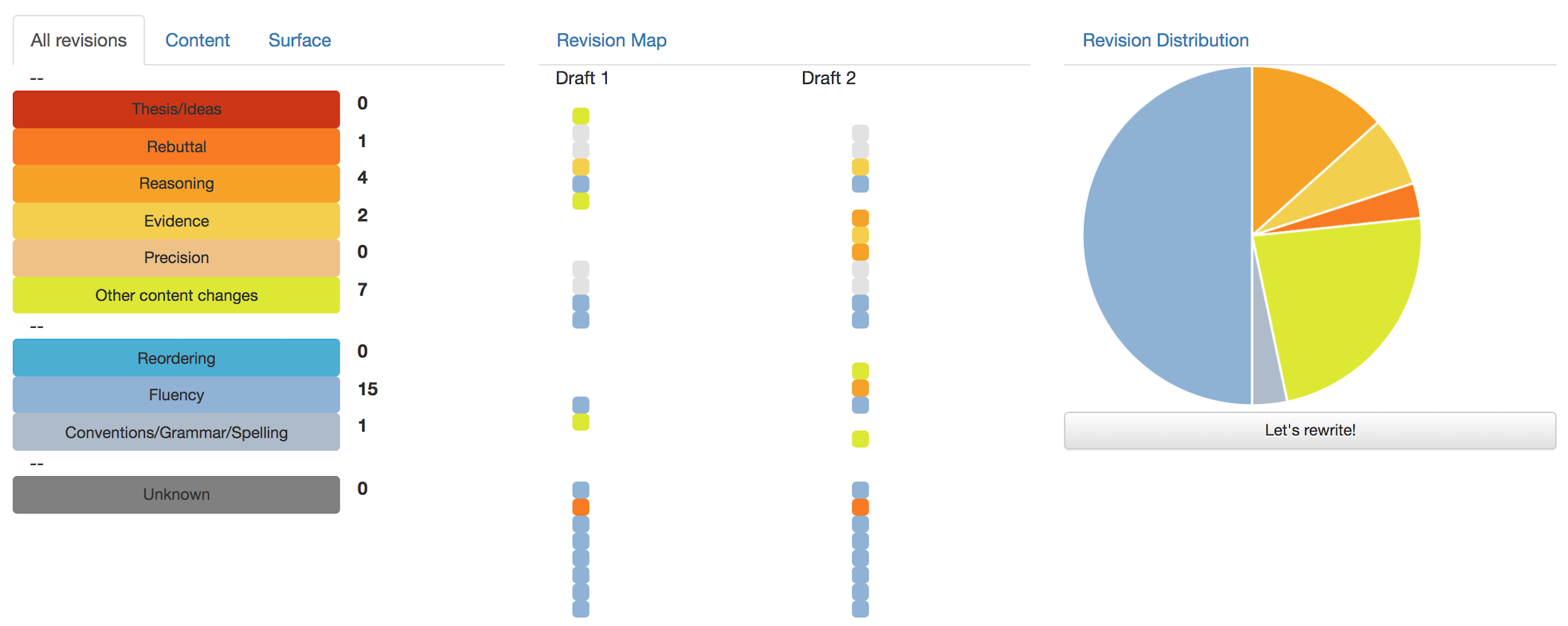}}
  \caption{Example of the overview interface from ArgRewrite condition C}
  \label{fig: overview interface}
  \Description{Example of the overview interface from ArgRewrite condition C.}
\end{figure}

    

The first interface that writers see after logging into ArgRewrite is the Overview interface. 
Here, writers are presented with overall visualizations of their revision patterns. 
The three main components of this overview interface are the revision purpose categories, the revision map, and the revision distribution pie chart. Figure~\ref{fig: overview interface} shows an example of the overview interface from ArgRewrite condition C.  
The revision purpose categories are highlighted with their corresponding colors on the left, the revision map is shown in the middle, and the revision distribution pie chart is shown on the right. The components are described below.
Once students are ready to revise their essay, they can click on the `Let's rewrite' button which leads them to the rewrite interface.

\subsubsection{Revision Purpose Categories}
\label{sec: Revision purpose categories}

\begin{figure}[h]
  
  \begin{subfigure}{0.5\linewidth}
  \centering
        \fbox{\includegraphics[width=0.83\linewidth]{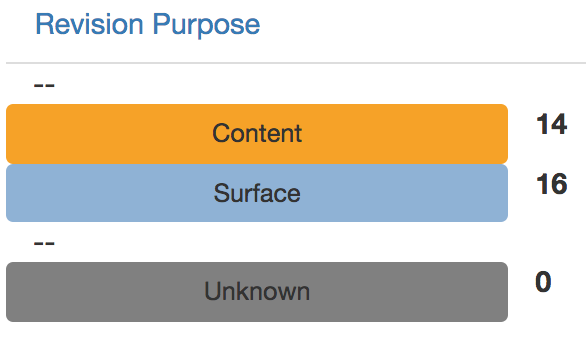}}
        \caption{Condition B}
        \label{fig:purpose-B}
        \Description{revision purpose categories for condition B}
    \end{subfigure}
    ~
  \begin{subfigure}{0.5\linewidth}
  \centering
        \fbox{\includegraphics[width=0.83\linewidth]{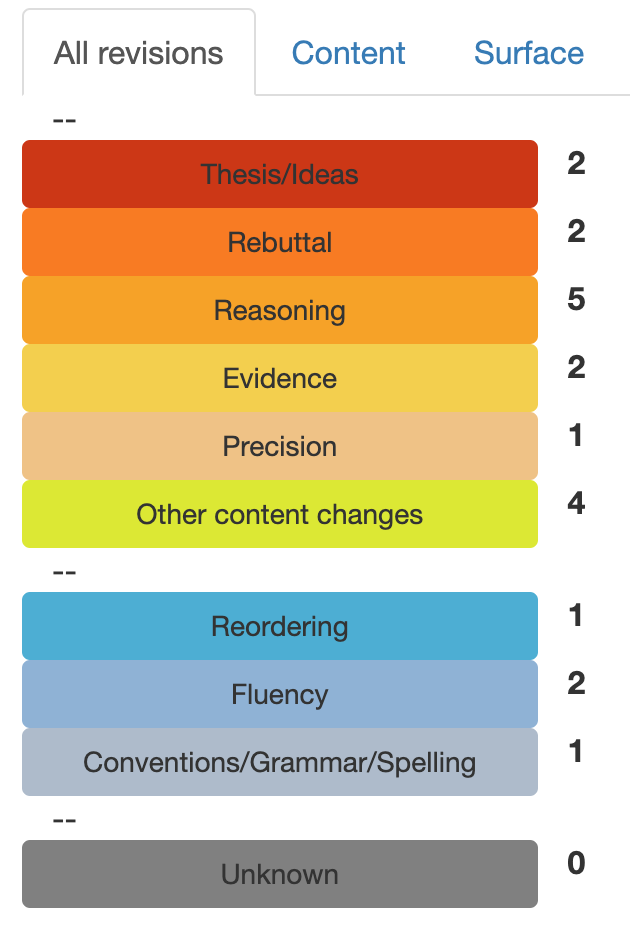}}
        \caption{Condition C,D}
        \label{fig:purpose-CD}
        \Description{revision purpose categories for condition C and D}
    \end{subfigure}
  \caption{Revision purpose categories}
  \label{fig: revision purpose categories}
  \Description{revision purpose categories}
\end{figure}

Based on the revision categories presented in ~\cite{zhang2015l}, our experiment addresses two principal categories of argumentative revisions -- surface and content. Surface revisions are the changes that do not alter the meaning of the sentence, e.g., convention or grammar, fluency, and organization changes. Content revisions consist of meaningful textual changes. Following previous works, we use six different categories of content changes -- claim, reasoning, evidence, rebuttal, precision, and other general changes\footnote{Precision category is added in addition to the content revisions reported in ~\cite{zhang2015l}.}. 
Figure~\ref{fig: revision purpose categories} shows the revision purpose categories for different conditions of the ArgRewrite interface. Following previous work~\cite{zhang2016hl}, surface and content revisions are shown in cold (e.g., blue) and warm (e.g., orange) colors, respectively. Condition B only shows binary revision categories, where the surface and content revisions are shown with blue and orange colors, respectively (shown in Figure~\ref{fig:purpose-B}). Figure~\ref{fig:purpose-CD} shows the detailed categories and the colors used for conditions C and D. Surface changes in conditions C and D are shown with different levels of blue colors from the cold color scale. Content changes are again shown with warm colors, but take up different colors from the warm color scale. If a revision does not fall into either of those categories, it is labeled as `unknown' and shown with gray color. The numbers in Figure~\ref{fig: revision purpose categories} represent the total added, deleted, and modified revisions for each revision category from Draft1 to Draft2.


\subsubsection{Revision Map}

\begin{figure}[h]
  \centering
  \begin{subfigure}{0.3\linewidth}
        \centering
        \includegraphics[width=0.35\linewidth]{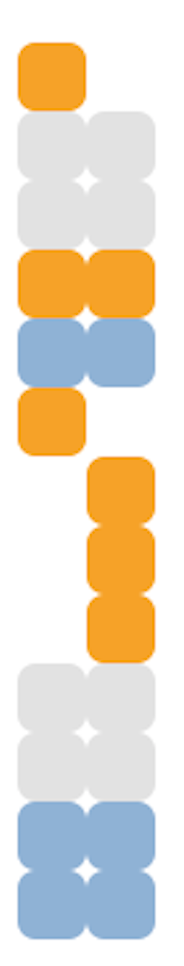}
        \caption{Condition B}
        \label{fig:rmap-B}
        \Description{revision map for condition B}
    \end{subfigure}
    ~
  \begin{subfigure}{0.3\linewidth}
        \centering        \includegraphics[width=0.34\linewidth]{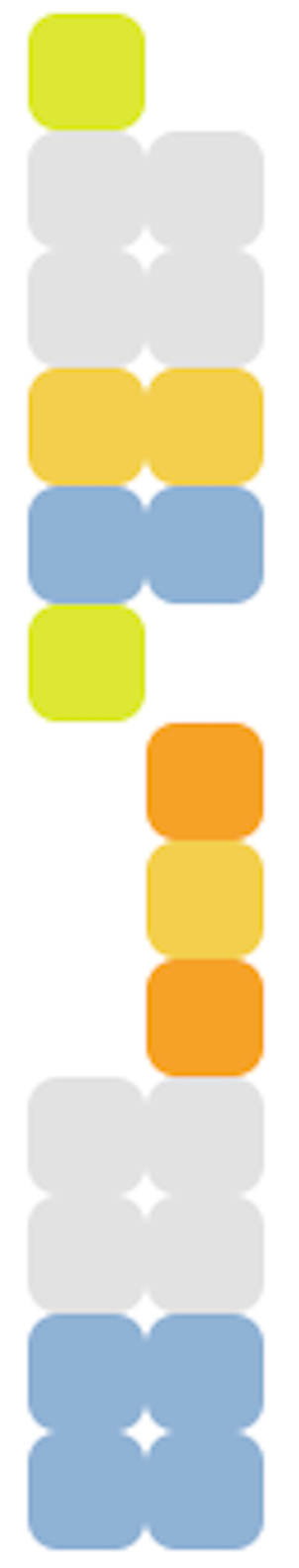}
        \caption{Condition C,D}
        \label{fig:rmap-CD}
        \Description{revision map for condition C and D}
    \end{subfigure}
  \caption{Revision map shows the drafts as two columns of color-coded tiles, each representing a sentence}
  \label{fig: revision map}
  \Description{Revision map for different conditions. Revision map shows the drafts as two columns of color-coded tiles, each representing a sentence.}
\end{figure}

Inspired by previous works ~\cite{southavilay2013revisionmap,zhang2016hl}, we design the revision map as two columns of aligned square tiles -- the left column represents the previous draft and the right column represents the current draft. Each tile represents a sentence in the draft; the white space between groups of tiles represents the paragraph breaks. Tiles are highlighted with colors of their corresponding revision categories. The shading of the tiles in each row represents whether the student added to, deleted, or modified the original sentence (or made no change). This revision map allows a student to look at all the revisions they made at different locations in the essay at a glance. Students can also easily understand what types of revisions they are making from the highlights. Figure~\ref{fig: revision map} shows the revision map for conditions B, C, and D. In Figure~\ref{fig:rmap-B}, the first tile is a deleted sentence because there is no aligned tile/sentence from the current draft. The orange color means it is a content revision. The light gray shade in the next two rows indicates that those sentences are not revised. Tiles in row 4 and 5 indicate modified content and surface revisions respectively. In contrast to the binary categories, Figure~\ref{fig:rmap-CD} shows the same revisions with fine-grained revision categories. It shows that the first sentence is a deleted general content revision, the fourth sentence is modified evidence, and the fifth sentence is a modified fluency revision.

\subsubsection{Revision Pie Chart}

\begin{figure}[h]
  \begin{subfigure}{0.5\linewidth}
  \centering
        \includegraphics[width=0.6\linewidth]{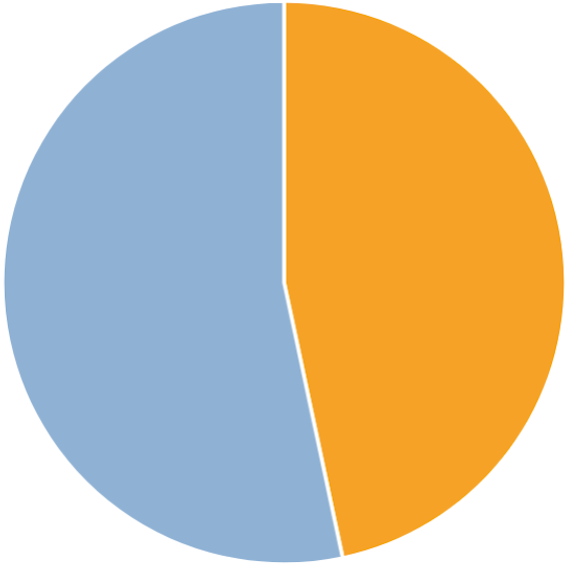}
        \caption{Condition B}
        \label{fig:rchart-B}
        \Description{revision chart for condition B}
    \end{subfigure}
    ~
  \begin{subfigure}{0.5\linewidth}
  \centering
        \includegraphics[width=0.6\linewidth]{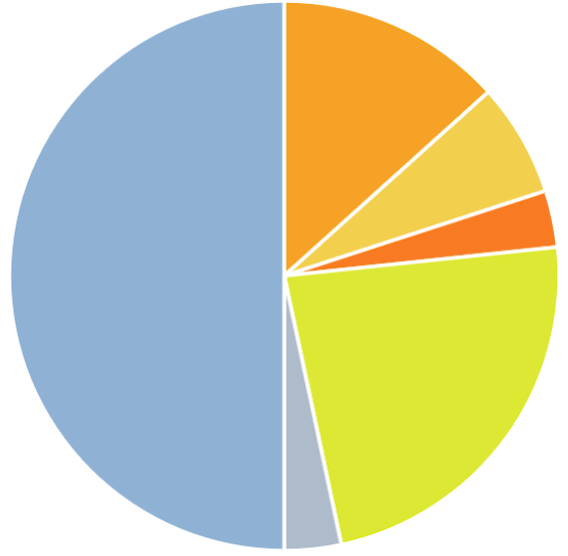}
        \caption{Condition C,D}
        \label{fig:rchart-CD}
        \Description{revision chart for condition C and D}
    \end{subfigure}
  \caption{Revision distribution shown as a pie-chart}
  \label{fig: revision chart}
  \Description{revision chart for different conditions}
\end{figure}

The overview interface also contains a pie chart showing the distribution of the frequency of different revision purpose categories.
While the revision purpose categories and the revision map show the number of revisions and the places where revisions are made, the pie chart adds the benefit of easy comparison of the distribution of different types of revisions. Looking at the pie chart, a student can easily understand the influence of the types of revisions they have made between Draft1 to Draft2. Figure~\ref{fig: revision chart} shows the revision chart from ArgRewrite conditions B, C, and D. Since we have only two revision types in condition B, Figure~\ref{fig:rchart-B} shows the distribution of the number of content and surface revisions. This chart (Figure~\ref{fig:rchart-B}) shows that this student made more surface than content revisions. Figure~\ref{fig:rchart-CD} shows similar information but provides additional details, such as the surface changes were predominately fluency changes, few grammar changes, while the main content changes involved reasoning and other (non-argumentative) content revisions.

\subsection{Rewrite Interface}

\begin{figure}[h]
  \begin{subfigure}{1.0\linewidth}
  \centering
        \includegraphics[width=1.0\linewidth]{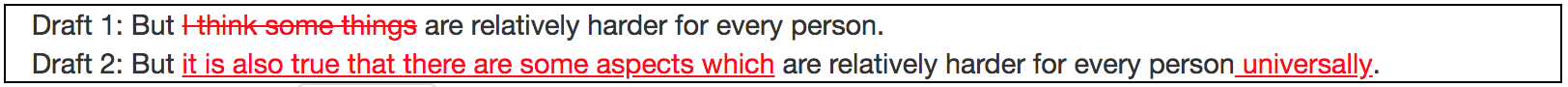}
        \caption{Condition B,C}
        \label{fig:rdetails-BC}
        \Description{revision detail window for condition B}
    \end{subfigure}
  \begin{subfigure}{1.0\linewidth}
  \centering
        \includegraphics[width=1.0\linewidth]{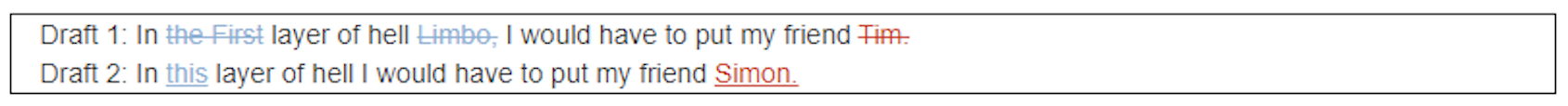}
        \caption{Condition D}
        \label{fig:rdetails-D}
        \Description{revision detail window for condition C and D}
    \end{subfigure}
  \caption{Revision details window for different conditions.}
  \label{fig: revision details window}
  \Description{Revision details window for different conditions}
\end{figure}





The rewrite interface contains the revision purpose categories, revision details window, four tabs containing the prompt and three essay drafts, and the revision map similar to the overview interface (except for condition D). Figures~\ref{fig: overview and rewrite interface B}, ~\ref{fig: overview and rewrite interface C}, and ~\ref{fig: overview and rewrite interface D} show screenshots of the rewrite interface for different conditions of the ArgRewrite. To encourage students the texts on the drafts tabs are highlighted with the corresponding revision color. In conditions B and C, the full sentence is highlighted. In condition D, only the revised text within a sentence is highlighted. Students can directly modify the essay on the Draft3 tab, which initially contains Draft2 to start with. 
When a student clicks on the text to see the details, a small window pops up to show the character-level differences\footnote{google diff match-patch: https://github.com/google/diff-match-patch} of a selected original and revised sentence. The character differences are highlighted with red in condition B and C. Condition D shows similar differences, but in corresponding revision purpose colors as shown in Figure~\ref{fig: revision details window}. 

The rewrite interface also provides the revision map of sentences to facilitate the navigation through the essay. Students can click on a tile on the revision map on the rewrite interface to look at that particular sentence. However, this is provided for conditions B and C only. Condition D shows a revision map for sub-sentential revisions; it shows two rows of tiles (shown at the top of the Figure~\ref{fig: rewrite interface D}) and each tile represents a revised sub-sentential unit within the revised sentences. 
On the rewrite interface, the small round button beside each tile of the revision map is used to highlight the confirmed revision categories when the students go through their previous revisions and submit their agreement about the revision categories.



\section{ArgRewrite Conditions}
\label{sec: ArgRewrite Interfaces}


\subsection{Condition A: No Revision Categorization}

\begin{figure}[!h]
  \centering
    \fbox{\includegraphics[width=0.99\linewidth]{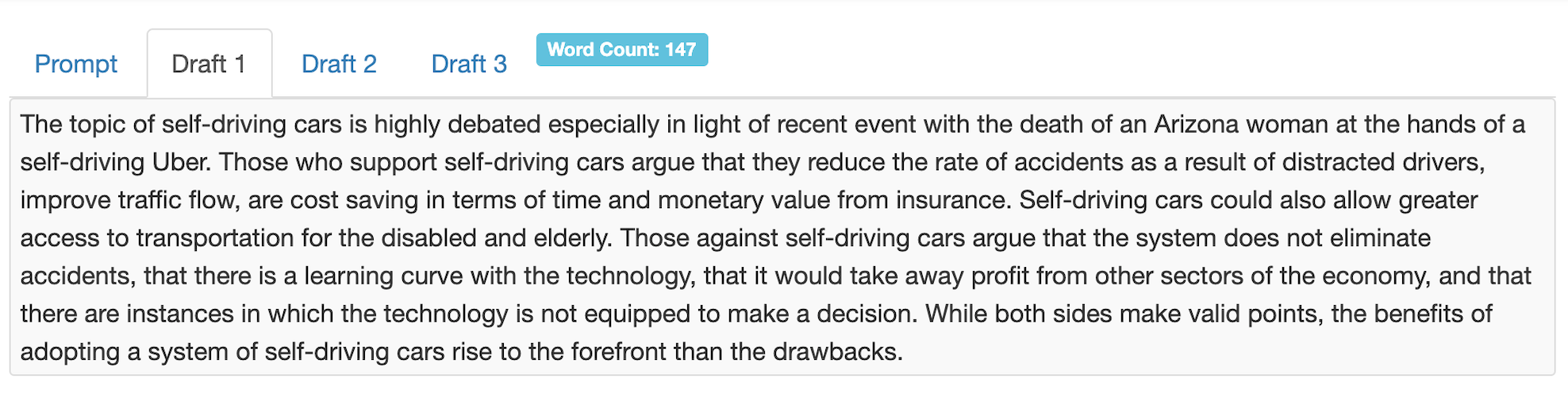}}
  \caption{A screenshot of the ArgRewrite tool - Condition A}
  \label{fig: cond A interface}
  \Description{ArgRewrite control interface}
\end{figure}

The ArgRewrite condition A is designed as a baseline containing no revision feedback, to compare with all other ArgRewrite conditions where writers receive different levels of feedback or analysis of their previous revision effort. Since there is no feedback, it does not contain any revision purpose categorization, revision map, or revision pie chart. Therefore, condition A does not have an overview interface. It contains a simplified version of the rewrite interface shown in Figure~\ref{fig: cond A interface}. The rewrite interface contains the plain text of the student essays for each Draft. 


\subsection{Condition B: Binary Revision Categorization}

\begin{figure}[!h]
  \centering
  \begin{subfigure}{0.92\linewidth}
        \fbox{\includegraphics[width=0.99\linewidth]{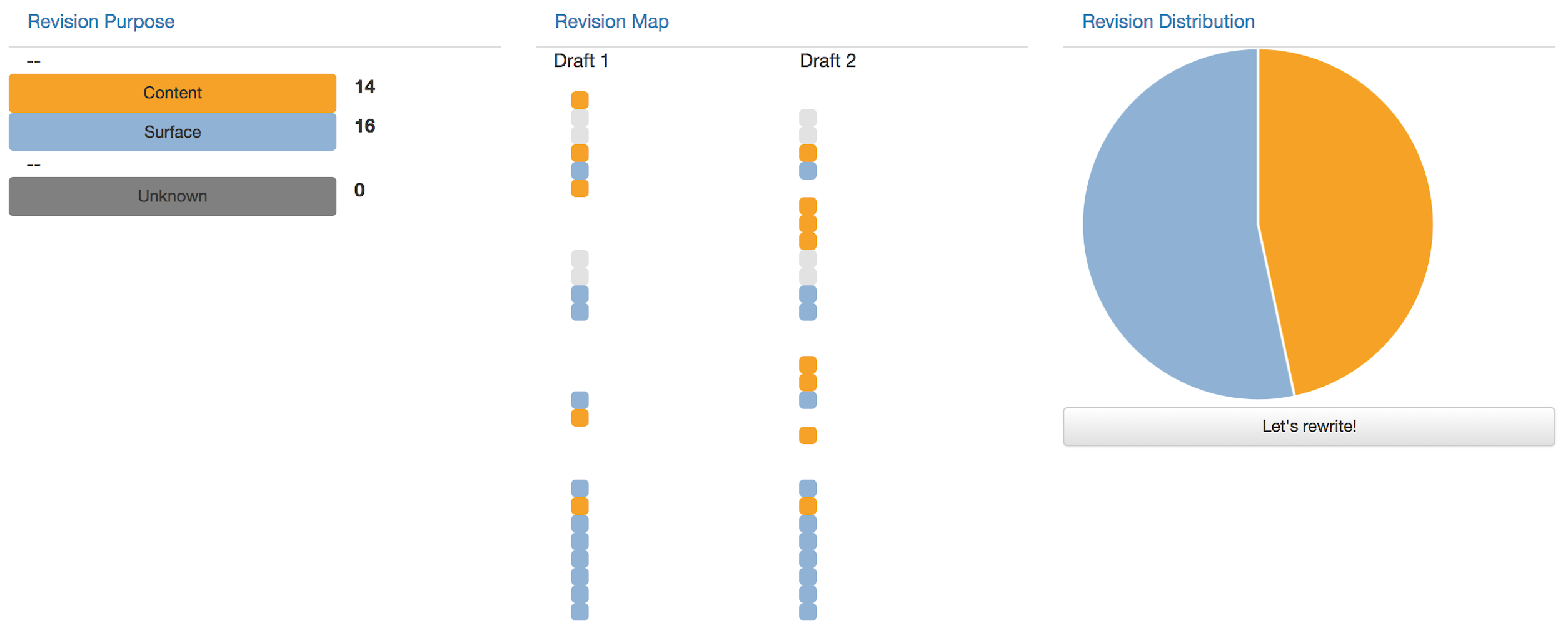}}
        \caption{Overview interface for condition B}
        \label{fig: overview interface B}
        \Description{Overview interface for condition B}
    \end{subfigure}
    
  \begin{subfigure}{0.92\linewidth}
        \fbox{\includegraphics[width=0.99\linewidth]{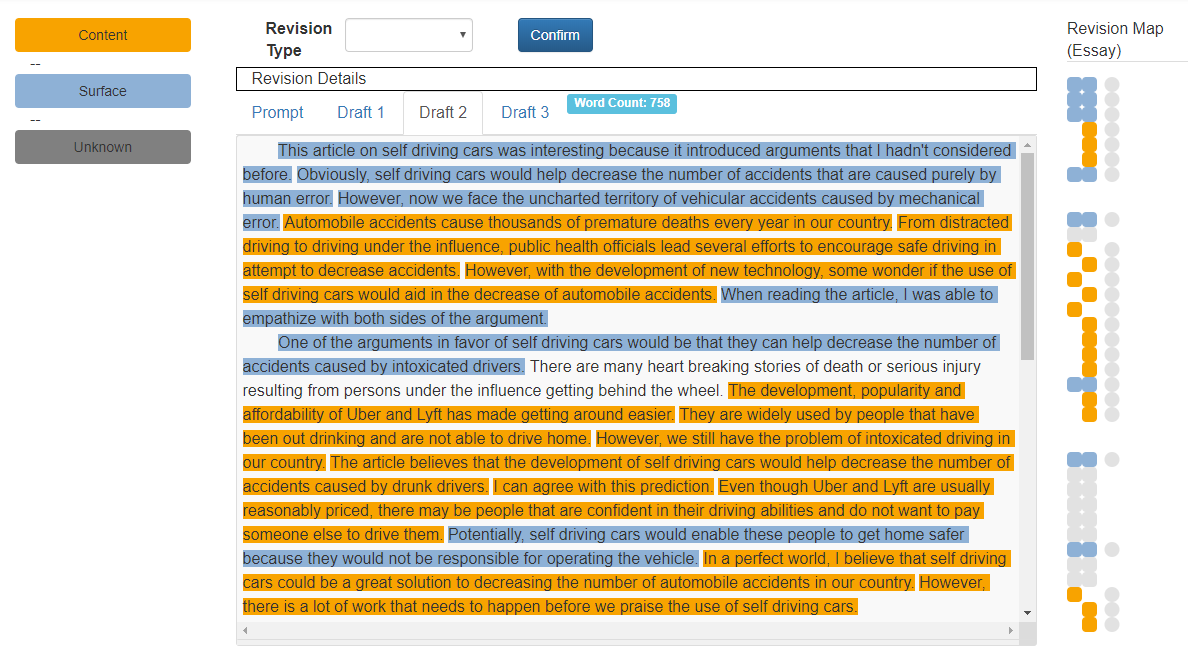}}
        \caption{Rewrite interface for condition B}
        \label{fig: rewrite interface B}
        \Description{Rewrite interface for condition B}
    \end{subfigure}
  \caption{A screenshot of the ArgRewrite tool - Condition B}
  \label{fig: overview and rewrite interface B}
  \Description{A screenshot of the ArgRewrite Interface (Condition B: Binary Revision Categorization).}
\end{figure}

ArgRewrite condition B is designed to provide simple revision feedback to the students. It includes all the components of the overview and the rewrite interface.  Revision categorization is shown at the sentence-level.
Condition B shows the revisions highlighted using only the top-level (binary) revision purpose categories - surface and content. The surface revisions are highlighted with blue and the content revisions are highlighted with orange to reflect cold versus warm color revisions as described in Section~\ref{sec: Revision purpose categories}. On the rewrite interface shown in Figure~\ref{fig: rewrite interface B}, if a sentence contains any surface revisions, the whole sentence is highlighted with blue. Similarly, sentences with content revisions are highlighted with orange. Similar to condition A, condition B also has four tabs to show the essay prompt and the drafts. Unlike conditions C and D, condition B is simple in terms of categorization of revisions. 

\subsection{Condition C: Detailed Revision Categorization}

\begin{figure}[h]
  \centering
  \begin{subfigure}{0.92\linewidth}
        \fbox{\includegraphics[width=0.99\linewidth]{Images/CondC-overview.png}}
        \caption{Overview interface for condition C}
        \label{fig: overview interface C}
        \Description{Overview interface for condition C}
    \end{subfigure}
    
  \begin{subfigure}{0.92\linewidth}
        \fbox{\includegraphics[width=0.99\linewidth]{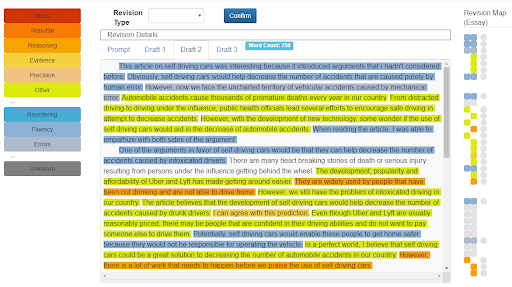}}
        \caption{Rewrite interface for condition C}
        \label{fig: rewrite interface C}
        \Description{Rewrite interface for condition C}
    \end{subfigure}
  \caption{A screenshot of the ArgRewrite tool - Condition C}
  \label{fig: overview and rewrite interface C}
  \Description{A screenshot of the ArgRewrite Interface (Condition C).}
\end{figure}

Condition C shows the detailed revision categorization, highlighted with their corresponding colors shown in Figure~\ref{fig: overview and rewrite interface C}. It contains all the components of the overview (Figure~\ref{fig: overview interface C}) and the rewrite interface (Figure~\ref{fig: rewrite interface C}). Students get the detailed revision feedback of their essay at sentence-level, according to the revision purpose categories described in Section~\ref{sec: Revision purpose categories}. In contrast to condition B, students who use condition C to revise their essay can, for example, spot the difference between word-usage versus grammar changes, claim versus evidence changes, etc. It is more informative 
compared to the control condition and to condition B with its binary revision categorization. Similarly to condition B, the rewrite interface in condition C also shows four tabs and highlights the whole sentence with the identified revision color.

\subsection{Condition D: Detailed Sub-Sentential Revision Categorization}

\begin{figure}[h]
  \centering
  \begin{subfigure}{0.92\linewidth}
        \fbox{\includegraphics[width=0.99\linewidth]{Images/CondC-overview.png}}
        \caption{Overview interface for condition D}
        \label{fig: overview interface D}
        \Description{Overview interface for condition D}
    \end{subfigure}

  \begin{subfigure}{0.92\linewidth}
        \fbox{\includegraphics[width=0.99\linewidth]{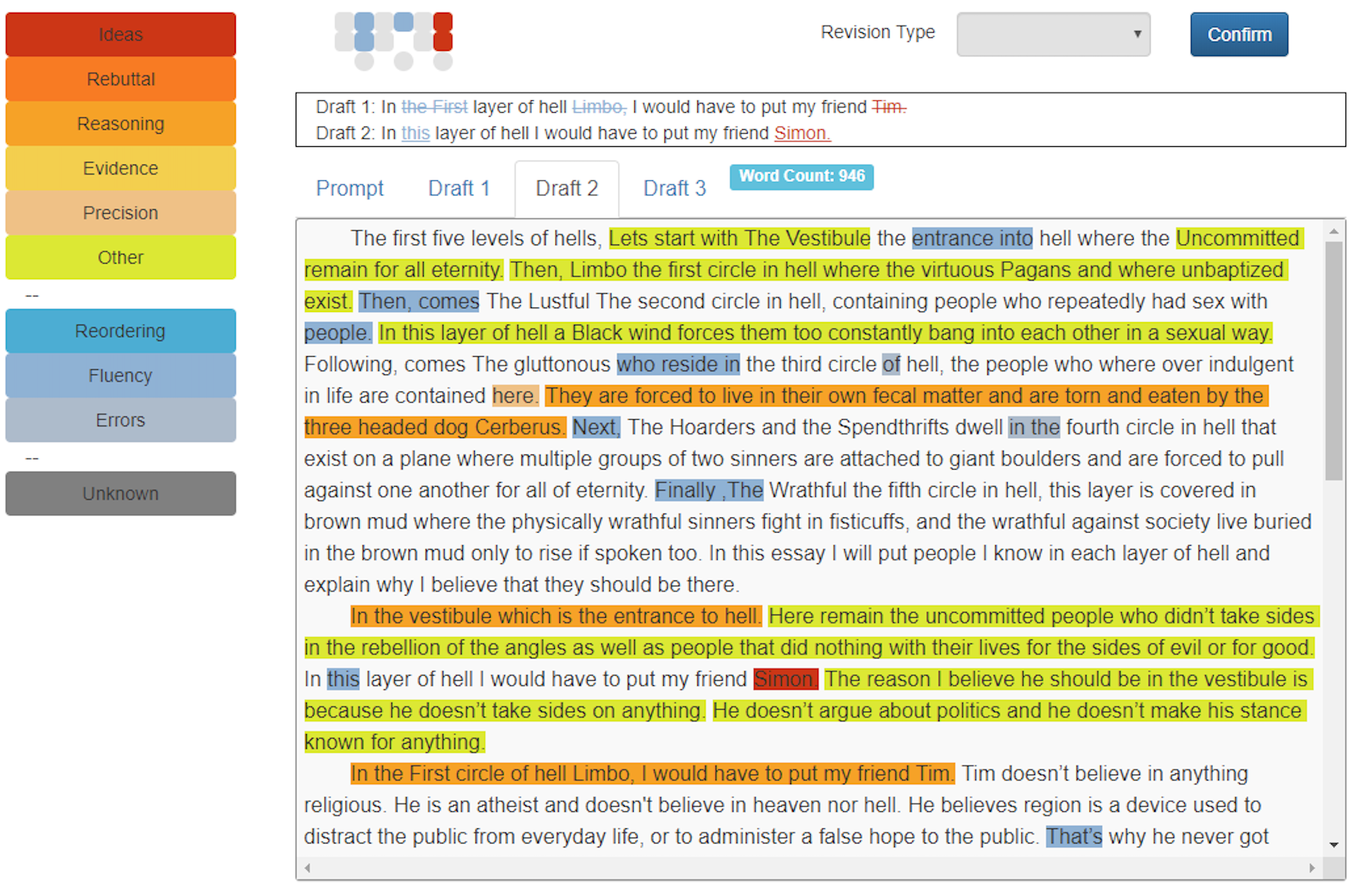}}
        \caption{Rewrite interface for condition D}
        \label{fig: rewrite interface D}
        \Description{Rewrite interface for condition D}
    \end{subfigure}
  \caption{A screenshot of the ArgRewrite tool- Condition D}
  \label{fig: overview and rewrite interface D}
  \Description{A screenshot of the ArgRewrite Interface (Condition D).}
\end{figure}


Condition D is designed to provide more detailed feedback for the revisions students make. Unlike conditions B and C, condition D can focus on multiple different revisions within a single sentence. Each sub-sentential revision is identified and highlighted with the corresponding revision category (shown in Figure~\ref{fig: rewrite interface D}). 
This condition contains an overview interface with a sentence-level revision map, similar to condition C, but the statistics of revision purpose categories are collected and shown from the sub-sentential revision units (Figure~\ref{fig:purpose-CD} and Figure~\ref{fig:rchart-CD}). 
In the rewrite interface, each sub-sentential revision is highlighted with its corresponding revision purpose color code. 
By clicking on each revised sentence, a horizontal revision map provides the abstract visualization of how it differs from the original sentence: which sub-sentential units are added, modified, or deleted, and what is the purpose of that revision.



\section{Evaluation and Results}

To evaluate our research hypothesis that more detailed feedback is more helpful (i.e., Condition D > C > B > A), we conducted an experiment to answer the following research questions.

\textbf{RQ1:} Do students perceive the ArgRewrite to be clear and easy to use?

\textbf{RQ2:} Do students find the ArgRewrite helpful for their writing? 


\textbf{RQ3:} Is ArgRewrite beneficial for student essay improvement?
 
\textbf{RQ4:} Is there any difference in students' revision behavior based on ArgRewrite condition?

\begin{table*}[!h]
  \caption{Interface survey questions, mean student response for each condition, and univariate ANOVA result with Fisher's least significant difference (LSD) procedure (*  p$<.05$,  **  p$<.01$, ***  p$<.001$, $\sim$  p$<.1$, $\alpha$=Cronbach's Alpha)}
  \label{tab: cond survey}
  \begin{tabular}{p{0.014\linewidth}p{0.35\linewidth}ccccp{0.16\linewidth}p{0.07\linewidth}p{0.08\linewidth}}
    \toprule									
 & 	&		&		&		&		& \multicolumn{3}{c}{ANOVA Result} \\
 & 	&	A	&	B	&	C	&	D	& Pairwise Comparison & F-score & Effect Size\\
\midrule
& Perceived ease of use ($\alpha$=0.87)	&	4.40 & 4.16 & 4.09 & 3.75	& D<A*  & 2.31$\sim$	&	0.08 \\
\midrule

1 & I find the system easy to use.	&	4.40	&	4.18	&	4.09	&	3.77	& D<A*
 & 1.69	&	0.06 \\
2 & My interaction with the system is clear and understandable.	&	4.40	&	4.14	&	4.09	&	3.73	& D<A**  & 2.48$\sim$	&	0.08 \\
\midrule
\midrule
& Perceived usefulness and usage behavior ($\alpha$=0.89)	&	3.53   & 3.77    & 4.33     & 4.11 & A<C***, A<D**, B<C**  & 6.06**	&	0.18 \\
\midrule
3 & The system allows me to have a better understanding of my previous revision efforts.	&	3.70	&	3.95	&	4.45	&	4.27	& A<C**, A<D*, B<C* & 3.40*	&	0.11 \\
4 & The system helps me to recognize the weakness of my essay.	&	3.10	&	3.32	&	4.09	&	3.73	& A<C**, A<D*, B<C* & 4.28**	&	0.14 \\
5 & Overall the system is helpful to my writing.	&	3.25	&	3.73	&	4.27	&	4.18	& A<C***, A<D***, B<C*  & 7.61***	&	0.22 \\
6 & The system encourages me to make more revisions (quantity) than I usually do.	&	3.65	&	3.86	&	4.50	&	4.09	& A<C**, B<C* & 3.28*	&	0.11 \\ 
7 & The system encourages me to make more meaningful revisions (quality) than I usually do.	&	3.45	&	3.86	&	4.23	&	4.23	& A<C*, A<D* & 2.95*	&	0.10 \\ 

8 & I put a lot of effort into writing and revising this essay.	&	4.00	&	3.91	&	4.41	&	4.14	& A<C*, B<C**  & 3.91*	&	0.13 \\
\midrule
\midrule
& Perceived usefulness and actual usage of the interface ($\alpha$=0.68)	&	--	&	4.05	&	4.15	&	4.04	& No difference  & 0.41	&	0.01 \\
\midrule									
9 & I found the overview page to be useful.	&	--	&	4.14	&	4.18	&	4.14	& No difference & 0.03	&	0.00 \\
10 & The description of the purpose of my revisions inspired me to make more revisions.	&	--	&	3.59	&	4.14	&	4.09	& B<C$\sim$ , B<D$\sim$ & 2.36	&	0.07 \\
11 & I  found it useful to see my revision purposes highlighted in different colors (i.e. Warm and cold colors)	&	--	&	4.27	&	4.64	&	4.41	& No difference & 1.75	&	0.05 \\
12 & I found the revision map visualization useful. 	&	--	&	4.09	&	3.86	&	3.73	& No difference & 0.72	&	0.02 \\
13 & I found the small window of revision details to be useful.	&	--	&	4.64	&	4.55	&	3.59	& D<B***, D<C***  & 13.73***	&	0.30 \\
14 & In general, I found it helpful to know whether my revision was a surface or content level change.	&	--	&	4.05	&	4.09	&	4.23	& No difference & 0.25	&	0.01 \\
15 & My revision purposes were most often indicated correctly by the system.	&	--	&	4.00	&	3.91	&	3.95	& No difference & 0.08	&	0.00 \\
16 & I trust the feedback that the system gave me.	&	--	&	3.59	&	3.86	&	4.18	& B < D*  & 2.94$\sim$	&	0.09 \\

\bottomrule									
  \end{tabular}
\end{table*}


Our analyses for RQ1 and RQ2 were based on data from a 16 question survey that participants completed after using ArgRewrite to revise their essays. The survey items addressed ~\cite{holden2011survey}'s distinction between "perceived ease of use" and "perceived usefulness" of technology. We included some questions verbatim from  ~\cite{holden2011survey}'s survey, such as questions 1 and 2, while other items were customized to address unique features of ArgRewrite
(shown in Table~\ref{tab: cond survey}).
Eight questions about the perceived ease of use and helpfulness and of the system for supporting essay revision were asked of all participants (questions 1-8). Another set of 8 questions (9-16) focused on usefulness of specific interface components and were asked only of participants in conditions B, C, and D. Each question was answered using a Likert scale ranging from 1 to 5 indicating strongly disagree to strongly agree. To answer RQ3 we examined students' writing improvement, based on expert essay scores that we describe below. Finally, we analyze the revision categories in student essays to answer RQ4. In our analyses, univariate analysis of variance (ANOVA) multiple comparison using Fisher's Least Significant Difference (LSD) test was used to compare differences in survey answers, essay scores, and number of revisions across different conditions. We calculate Cronbach's Alpha coefficient to report internal consistency of the combined survey questions (shown in Table~\ref{tab: cond survey}). In RQ4, we also use t-test to compare revisions within conditions.


To answer \textbf{RQ1}, we combine two survey questions (1-2) that ask about the perceived ease of use of the tool. The questions asked students if they find the system easy to use, and if their interaction with the system is clear and understandable.  Mean survey ratings and ANOVA result for those questions are shown in Table~\ref{tab: cond survey}. For perceived ease of use, the overall difference between conditions is not significant. Looking at pairwise comparison, 
condition A has a higher mean compared to all other conditions, and Condition D has the lowest mean. Condition A, which is the control condition without any revision feedback, was thus the easiest condition to use. This is not surprising because of the simplicity of the rewrite interface for condition A. However, this mean-value is only significantly higher than condition D, where we provided the most specific revision feedback. We think this lower mean value reflects the complex information display of the revision categories at the sub-sentence level. 

To answer \textbf{RQ2}, we first combine the survey questions  
(3--8) that focus on the perceived usefulness~\cite{holden2011survey} and usage behavior. We then separately examine questions (9--16) regarding usefulness and actual usage of the interface components. Taking the means over questions 3--8 shows that overall, there is a significant difference between conditions although the ANOVA effect size is low. Students perceived condition C with detailed sentence-level revision feedback to be more useful compared to conditions A and B. 
Particularly, ANOVA results from Table~\ref{tab: cond survey} shows that students using condition C thought that the system helps them to better understand their previous revision effort and recognize their weakness, encourages to make more revisions, and more helpful compared to students using conditions A and B. In other words, from this ANOVA result we can say that condition A proved to be less helpful (despite being the easiest to use).
Students also perceived detailed sub-sentential revision feedback to be more useful compared to no feedback. 
For example, when we asked about the quality of revision\footnote{Students received instruction in the tutorial that content revisions were more related to essay improvement in previous studies~\cite{zhang2015l}. They were encouraged to do more content revisions.}, condition D showed a significantly higher mean-value than condition A. 
Overall, we can say that detailed feedback is more useful than no feedback or binary feedback which supports our hypothesis.  
However, we did not see any significant difference between sentence versus sub-sentential revision feedback (C versus D). Therefore we speculate that reducing the granularity of revision feedback might not be very beneficial after all. 

We get a mixed signal looking at the questions (9--16) that only target the conditions with feedback (B, C, and D). Overall, ANOVA shows no significant difference between conditions for this group of questions that focus on the actual usage of the interface. However, pairwise comparisons do show some significant differences. For example, students find the revision windows more helpful when they were shown sentence-level revision feedback compared to sub-sentential feedback. However, most of the specific components of the overview and rewrite interface did not show any difference between the conditions (e.g., revision map). 
On the other hand, a detailed description of revision purpose seemed more inspiring than the binary description (question 10). 
Detailed sub-sentential feedback was also trustworthy compared to sentence-level binary feedback.  Given the Wizard of Oz scenario, the accuracy of the system feedback is objectively similar across conditions.

\begin{table*}
    \vspace*{10mm}
  \caption{Average score and normalized essay score gain (NEG) per condition (*  p$<.05$, $\sim$  p$<.1$).}
  \label{tab:nlg}
  \begin{tabular}{lcccccccc}
    \toprule
	&		&		&		&		&		&	\multicolumn{3}{c}{ANOVA Result}	\\
	&		&	A	&	B	&	C	&	D	&	Pairwise comparison			&	F-score	&	Effect Size\\
	\midrule
\multirow{3}{*}{Score in}	&	Draft1	&	23.38	&	24.15	&	22.80	&	24.23	& \multirow{3}{*}{-}				&	\multirow{3}{*}{-}	& \multirow{3}{*}{-}	\\
	&	Draft2	&	23.90	&	25.64	&	24.55	&	25.25	&		&		&		\\
	&	Draft3	&	24.90	&	26.50	&	25.84	&	26.11	&			&		&		\\
	\midrule
\multirow{2}{*}{NEG}	&	1 to 3	&	0.06	&	0.14	&	0.18	&	0.13	&		A<C* , A<B$\sim$		&	2.38$\sim$	& 0.08	\\ 
	&	2 to 3	&	0.05	&	0.07	&	0.08	&	0.06	&	No significant difference			&	0.55	& 0.02	\\
	
    
  \bottomrule
\end{tabular}
\end{table*}

To answer \textbf{RQ3}, we looked at students' essay score. 
All three drafts written by each participant were scored separately by two researchers, both of whom were experienced high school English and college instructors. The quadratic weighted kappa (QWK) is 0.537. Scoring was guided by a 10-criteria rubric that mirrored the rubric\footnote{The rubric is provided in \ref{a sec: rubric}} used to give feedback on Draft1 focusing on the argument elements in the essay. Each item was scored on a scale of 1-4: ``1-poor,''         ``2-developing,'' ``3-proficient,'' or ``4-excellent.'' 
The essay score ranges from 10 to 40.
The average of the two researchers' scores was used for data analysis.
To determine the improvement of student essay we calculated the normalized essay score gain (NEG) from Draft1 to Draft3 ($NEG13$) and Draft2 to Draft3 ($NEG23$). We did not consider the essay score gain from Draft1 to Draft2 because that step does not involve using our system. Normalized essay score gain is calculated as follows:

\[ NEG = \frac{Current Draft Score - Previous Draft Score}{MaxScore - Previous Draft Score} \]

For both $NEG13$ and $NEG23$, we have the highest mean-value for condition C, where we showed the detailed sentence-level revision feedback (Table~\ref{tab:nlg}). We again performed univariate ANOVA with Fisher LSD test to compare the mean of the essay score gains in different interface conditions. The overall ANOVA result did not show any significant difference. ANOVA pairwise comparison result for $NEG13$ showed that students in Condition C performed significantly better than condition A. Condition B was trending better than Condition A ($p = 0.06$). But there was no significant difference between B, C, and D. We also did not see any significant difference for $NEG23$ between any conditions. 
This result is in line with our previous research question results, in which we observed that students found detailed sentence-level revision feedback to be more helpful compared to no revision feedback at all.

\begin{table*}
  \caption{Number of sentence-level surface and content revisions between first (Draft1 to Draft2) and second (Draft2 to Draft3) revision stage for each condition. ($\uparrow$: increase in number of revisions compared to the previous revision stage, $\downarrow$: decrease in number of revisions compared to the previous revision stage, *  p$<.05$, $\sim$  p$<.1$)}
  \label{tab: revision stat}
  \begin{tabular}{clcccc}
    \toprule
    Revision Stage	&	Revision	&	A	&	B	&	C	&	D	\\
    \midrule
\multirow{2}{*}{1 to 2}	&	Surface	&	131 (30\%)	&	136 (37\%)	&	202 (47\%)	&	164 (41\%)	\\
	                    &	Content	&	322 (70\%)	&	235 (62\%)	&	239 (53\%)	&	230 (59\%)	\\
\midrule
\multirow{2}{*}{2 to 3}	&	Surface	&	185 (40\%) $\uparrow$*	&	160 (47\%) $\uparrow$*	&	198 (48\%) 	$\sim$	&	183 (45\%) 	$\sim$	\\
	                    &	Content	&	280 (60\%) $\downarrow$*	&	189 (53\%) $\downarrow$*	&	213 (52\%) 	$\sim$	&	225 (55\%) 	$\sim$	\\
  \bottomrule
\end{tabular}
\end{table*}

To answer \textbf{RQ4}, we looked at the types of revisions (surface vs. content) students made when revising Draft1 to Draft2 (without ArgRewrite) and when revising Draft2 to Draft3 (with ArgRewrite). We expected to see fewer revisions with ArgRewrite since it is the second stage of revising the same essay. 
Table~\ref{tab: revision stat} shows the percentage of surface and content revisions for each condition. Within each condition, we compare the number of surface and content revisions across revision stage using paired t-test. 
In conditions A and B, we observed significantly more surface revisions and fewer content revisions when revising using ArgRewrite compared to revising without ArgRewrite, but the distribution of types of revisions is not significantly different in condition C and D, when with or without ArgRewrite.

ANOVA result showed no significant difference between conditions for the average number of content or surface revisions. As we have mentioned before, according to previous work, content revisions (e.g. reasoning, evidence) are correlated with essay improvement. Hence, according to Table~\ref{tab: revision stat}, students in condition A should have higher essay score gains with more content revisions than others. But in Table~\ref{tab:nlg} we have seen that condition A has the lowest essay score gain. With the lowest percentage of content revisions in condition C, students in that condition had higher essay score gains. This result indicates that students who received revision feedback generated revisions that help them improve the essay compared to students who did not receive any feedback. Although students with no feedback generated more content revisions, we speculate those revisions may be irrelevant or unnecessary for supporting the argument.


\section{Discussion}


The findings of this study highlight a tension point that is worth further examination.
On the one hand, the analysis of the improvement and revision patterns suggested that Condition C's detailed categorization of revision functions was more effective and helpful than the other conditions. On the other hand, there was an inverse relationship between the granularity of feedback and the usability of the system. In other words, the more detailed the feedback was on students' revision habits, the less students were likely to find it ``easy to use'' or ``clear and understandable'' (see questions 1 and 2 on Table~\ref{tab: cond survey}).  
Our findings consistently showed that feedback on detailed revision categorization is better than no feedback. For some evaluation measures, detailed feedback is also better than binary feedback. However, we did not find much difference between sentence versus sub-sentence level revision feedback. So our hypothesis that the more detailed the revision feedback the better is not entirely supported. One potential confound in our study design may have been the different units of analysis employed in Condition D versus the other conditions. By being provided  with sub-sentential as opposed to sentential feedback, writers in Condition D spent more time confirming the accuracy of their previous revisions than others. This resulted in them 
spending more time to look at previous revisions and 
less time to engage in the actual act of revising when it came to developing their last drafts. This likely contributed to their lower ratings of perceived ease of use, but it also may have influenced the quality of their final drafts. 
With this in mind, our analyses found little difference between conditions C and D. In the future, we plan to look at the sub-sentence level revisions more closely to understand how to make it more effective for the students. For example, we did not test binary revision categorization at the sub-sentence-level. This is a future condition we would like to explore.
Another significant difference we find between sentence-level and sub-sentential interface components is the small window of revision details. Students using sentence-level revision conditions find it more useful than students using sub-sentential revision feedback. We have seen before that the revision details window is different for condition D. It shows the sub-sentence revisions highlighted. So in condition D, students look at the sub-sentential highlights on the essay text and the revision details window, which is redundant. This might be the reason why the revision window was not good enough for condition D but showed to be very useful for conditions B and C.

On one final note regarding our third question related to student improvement, our analyses of improvement from first to third drafts seems to favor detailed sentence-level revision categorization.
In our study students revised their Draft1 at home. Hence, the revision from the first to second draft did not involve ArgRewrite. When students used our tool from the second to third draft, they still saw higher essay score gain using sentence-level revision feedback (binary and detailed) than sub-sentential, but those differences were not statistically significant. This might suggest that sub-sentential revision feedback is not helping students improve the essay, even compared to no revision feedback. However, due to the necessary methodological differences mentioned above,we believe we still need to conduct more experiments with sub-sentential revision before reaching any conclusion.



\section{Conclusion}
In this paper, we presented a tool that helps students to make further revisions on their argumentative writings. We developed four versions of the interface for the tool and presented a comparative study to determine to what extent might the explicit representations of revision purpose categories help students to improve their essay. Our analysis shows that detailed revision categorization at the sentence-level is the most helpful compared to conditions that do not provide detailed feedback. Detailed sub-sentential revision categorization also seemed promising, but more research and development is warranted. In particular, determining the most useful and intuitive level of granularity and detail in writing feedback is an open research question. In the future, we plan to further explore the sub-sentential revision purpose taxonomy 
to support effective automated writing assistant systems.
\begin{acks}
We would like to acknowledge Meghan Dale and Sonia Cromp for their work on this study. Special thanks to Dr. Erin Walker for her valuable suggestions on an early stage of this article. We would also like to thank the anonymous reviewers for taking the time to review our paper and provide us with detailed feedback. This work is supported by National Science Foundation (NSF) grant 1735752 to the University of Pittsburgh. The opinions expressed are those of the authors and do not represent the views of the Institute. 
\end{acks}

\bibliographystyle{ACM-Reference-Format}
\bibliography{references}


\begin{thebibliography}{35}


\ifx \showCODEN    \undefined \def \showCODEN     #1{\unskip}     \fi
\ifx \showDOI      \undefined \def \showDOI       #1{#1}\fi
\ifx \showISBNx    \undefined \def \showISBNx     #1{\unskip}     \fi
\ifx \showISBNxiii \undefined \def \showISBNxiii  #1{\unskip}     \fi
\ifx \showISSN     \undefined \def \showISSN      #1{\unskip}     \fi
\ifx \showLCCN     \undefined \def \showLCCN      #1{\unskip}     \fi
\ifx \shownote     \undefined \def \shownote      #1{#1}          \fi
\ifx \showarticletitle \undefined \def \showarticletitle #1{#1}   \fi
\ifx \showURL      \undefined \def \showURL       {\relax}        \fi
\providecommand\bibfield[2]{#2}
\providecommand\bibinfo[2]{#2}
\providecommand\natexlab[1]{#1}
\providecommand\showeprint[2][]{arXiv:#2}

\bibitem[\protect\citeauthoryear{Attali and Burstein}{Attali and
  Burstein}{2006}]%
        {attali2006b}
\bibfield{author}{\bibinfo{person}{Yigal Attali} {and} \bibinfo{person}{Jill
  Burstein}.} \bibinfo{year}{2006}\natexlab{}.
\newblock \showarticletitle{The Automated Essay Scoring with E-Rater V.2}.
\newblock \bibinfo{journal}{\emph{Journal of Technology, Learning, and
  Assessment}} \bibinfo{volume}{4}, \bibinfo{number}{3} (\bibinfo{year}{2006}).
\newblock


\bibitem[\protect\citeauthoryear{Bronner and Monz}{Bronner and Monz}{2012}]%
        {bronner2012m}
\bibfield{author}{\bibinfo{person}{Amit Bronner} {and}
  \bibinfo{person}{Christof Monz}.} \bibinfo{year}{2012}\natexlab{}.
\newblock \showarticletitle{User Edits Classification Using Document Revision
  Histories}. In \bibinfo{booktitle}{\emph{Proceedings of the 13th Conference
  of the European Chapter of the Association for Computational Linguistics}}
  \emph{(\bibinfo{series}{EACL '12})}. \bibinfo{publisher}{Association for
  Computational Linguistics}, \bibinfo{address}{Avignon, France},
  \bibinfo{pages}{356--366}.
\newblock
\showISBNx{978-1-937284-19-0}


\bibitem[\protect\citeauthoryear{Browne}{Browne}{2019}]%
        {browne2019chiWOZ}
\bibfield{author}{\bibinfo{person}{Jacob~T. Browne}.}
  \bibinfo{year}{2019}\natexlab{}.
\newblock \showarticletitle{Wizard of Oz Prototyping for Machine Learning
  Experiences}. In \bibinfo{booktitle}{\emph{Extended Abstracts of the 2019 CHI
  Conference on Human Factors in Computing Systems}} (Glasgow, Scotland Uk)
  \emph{(\bibinfo{series}{CHI EA ’19})}. \bibinfo{publisher}{Association for
  Computing Machinery}, \bibinfo{address}{New York, NY, USA},
  \bibinfo{pages}{1–6}.
\newblock
\showISBNx{9781450359719}


\bibitem[\protect\citeauthoryear{Chernodub, Oliynyk, Heidenreich, Bondarenko,
  Hagen, Biemann, and Panchenko}{Chernodub et~al\mbox{.}}{2019}]%
        {chernodub2019targer}
\bibfield{author}{\bibinfo{person}{Artem Chernodub}, \bibinfo{person}{Oleksiy
  Oliynyk}, \bibinfo{person}{Philipp Heidenreich}, \bibinfo{person}{Alexander
  Bondarenko}, \bibinfo{person}{Matthias Hagen}, \bibinfo{person}{Chris
  Biemann}, {and} \bibinfo{person}{Alexander Panchenko}.}
  \bibinfo{year}{2019}\natexlab{}.
\newblock \showarticletitle{TARGER: Neural Argument Mining at Your Fingertips}.
  In \bibinfo{booktitle}{\emph{Proceedings of the 57th Annual Meeting of the
  Association of Computational Linguistics (ACL'2019)}}.
  \bibinfo{address}{Florence, Italy}.
\newblock


\bibitem[\protect\citeauthoryear{Daxenberger and Gurevych}{Daxenberger and
  Gurevych}{2012}]%
        {daxenberger2012g}
\bibfield{author}{\bibinfo{person}{Johannes Daxenberger} {and}
  \bibinfo{person}{Iryna Gurevych}.} \bibinfo{year}{2012}\natexlab{}.
\newblock \showarticletitle{A Corpus-Based Study of Edit Categories in Featured
  and Non-Featured Wikipedia Articles}. In
  \bibinfo{booktitle}{\emph{Proceedings of the 24th International Conference on
  Computational Linguistics}} \emph{(\bibinfo{series}{COLING '12})}.
  \bibinfo{address}{Mumbai, India}, \bibinfo{pages}{711--726}.
\newblock


\bibitem[\protect\citeauthoryear{Eli~Review}{Eli~Review}{2014}]%
        {elireview}
\bibfield{author}{\bibinfo{person}{The Eli~Review}.}
  \bibinfo{year}{2014}\natexlab{}.
\newblock \bibinfo{title}{https://elireview.com. [Online; accessed
  01-12-2021].}
\newblock
\newblock


\bibitem[\protect\citeauthoryear{Fitzgerald and Shanahan}{Fitzgerald and
  Shanahan}{2000}]%
        {fitzgerald2000reading}
\bibfield{author}{\bibinfo{person}{Jill Fitzgerald} {and}
  \bibinfo{person}{Timothy Shanahan}.} \bibinfo{year}{2000}\natexlab{}.
\newblock \showarticletitle{Reading and writing relations and their
  development}.
\newblock \bibinfo{journal}{\emph{Educational Psychologist}}
  \bibinfo{volume}{35}, \bibinfo{number}{1} (\bibinfo{year}{2000}),
  \bibinfo{pages}{39--50}.
\newblock


\bibitem[\protect\citeauthoryear{Grammarly}{Grammarly}{2016}]%
        {grammarly}
\bibfield{author}{\bibinfo{person}{Grammarly}.}
  \bibinfo{year}{2016}\natexlab{}.
\newblock \bibinfo{title}{http://www.grammarly.com. [Online; accessed
  01-12-2021].}
\newblock
\newblock


\bibitem[\protect\citeauthoryear{Holden and Rada}{Holden and Rada}{2011}]%
        {holden2011survey}
\bibfield{author}{\bibinfo{person}{Heather Holden} {and} \bibinfo{person}{Roy
  Rada}.} \bibinfo{year}{2011}\natexlab{}.
\newblock \showarticletitle{Understanding the influence of perceived usability
  and technology self-efficacy on teachers’ technology acceptance}.
\newblock \bibinfo{journal}{\emph{Journal of Research on Technology in
  Education}} \bibinfo{volume}{43}, \bibinfo{number}{4} (\bibinfo{year}{2011}),
  \bibinfo{pages}{343--367}.
\newblock


\bibitem[\protect\citeauthoryear{Jones}{Jones}{2008}]%
        {johnes2008wikirevision}
\bibfield{author}{\bibinfo{person}{John Jones}.}
  \bibinfo{year}{2008}\natexlab{}.
\newblock \showarticletitle{Patterns of Revision in Online Writing: A Study of
  Wikipedia's Featured Articles}.
\newblock \bibinfo{journal}{\emph{Written Communication}} \bibinfo{volume}{25},
  \bibinfo{number}{2} (\bibinfo{year}{2008}), \bibinfo{pages}{262--289}.
\newblock


\bibitem[\protect\citeauthoryear{Knight, Shibani, Abel, Gibson, Ryan, Sutton,
  Wight, Lucas, Sandor, Kitto, et~al\mbox{.}}{Knight et~al\mbox{.}}{2020}]%
        {knight2020acawriter}
\bibfield{author}{\bibinfo{person}{Simon Knight}, \bibinfo{person}{Antonette
  Shibani}, \bibinfo{person}{Sophie Abel}, \bibinfo{person}{Andrew Gibson},
  \bibinfo{person}{Philippa Ryan}, \bibinfo{person}{Nicole Sutton},
  \bibinfo{person}{Raechel Wight}, \bibinfo{person}{Cherie Lucas},
  \bibinfo{person}{Agnes Sandor}, \bibinfo{person}{Kirsty Kitto},
  {et~al\mbox{.}}} \bibinfo{year}{2020}\natexlab{}.
\newblock \showarticletitle{AcaWriter: A learning analytics tool for formative
  feedback on academic writing}.
\newblock \bibinfo{journal}{\emph{Journal of Writing Research}}
  \bibinfo{volume}{12}, \bibinfo{number}{1} (\bibinfo{year}{2020}),
  \bibinfo{pages}{299--344}.
\newblock


\bibitem[\protect\citeauthoryear{Liu, Calvo, and Pardo}{Liu
  et~al\mbox{.}}{2013}]%
        {liu2013Tracer-tool}
\bibfield{author}{\bibinfo{person}{Ming Liu}, \bibinfo{person}{Rafael~A.
  Calvo}, {and} \bibinfo{person}{Abelardo Pardo}.}
  \bibinfo{year}{2013}\natexlab{}.
\newblock \showarticletitle{Tracer: A Tool to Measure and Visualize Student
  Engagement in Writing Activities}. In \bibinfo{booktitle}{\emph{Proceedings
  of the 2013 IEEE 13th International Conference on Advanced Learning
  Technologies}} \emph{(\bibinfo{series}{ICALT '13})}. \bibinfo{publisher}{IEEE
  Computer Society}, \bibinfo{address}{USA}, \bibinfo{pages}{421–425}.
\newblock


\bibitem[\protect\citeauthoryear{Loretto, DeMartino, and Godley}{Loretto
  et~al\mbox{.}}{2016}]%
        {loretto2016secondary}
\bibfield{author}{\bibinfo{person}{Adam Loretto}, \bibinfo{person}{Sara
  DeMartino}, {and} \bibinfo{person}{Amanda Godley}.}
  \bibinfo{year}{2016}\natexlab{}.
\newblock \showarticletitle{Secondary students' perceptions of peer review of
  writing}.
\newblock \bibinfo{journal}{\emph{Research in the Teaching of English}}
  (\bibinfo{year}{2016}), \bibinfo{pages}{134--161}.
\newblock


\bibitem[\protect\citeauthoryear{MacArthur, Philippakos, and Ianetta}{MacArthur
  et~al\mbox{.}}{2015}]%
        {macarthur2015self}
\bibfield{author}{\bibinfo{person}{Charles~A MacArthur}, \bibinfo{person}{Zoi~A
  Philippakos}, {and} \bibinfo{person}{Melissa Ianetta}.}
  \bibinfo{year}{2015}\natexlab{}.
\newblock \showarticletitle{Self-regulated strategy instruction in college
  developmental writing.}
\newblock \bibinfo{journal}{\emph{Journal of Educational Psychology}}
  \bibinfo{volume}{107}, \bibinfo{number}{3} (\bibinfo{year}{2015}),
  \bibinfo{pages}{855}.
\newblock


\bibitem[\protect\citeauthoryear{Newell, VanDerHeide, and Olsen}{Newell
  et~al\mbox{.}}{2014}]%
        {newell2014HighSE}
\bibfield{author}{\bibinfo{person}{George~E. Newell}, \bibinfo{person}{Jennifer
  VanDerHeide}, {and} \bibinfo{person}{Allison~Wynhoff Olsen}.}
  \bibinfo{year}{2014}\natexlab{}.
\newblock \showarticletitle{High School English Language Arts Teachers'
  Argumentative Epistemologies for Teaching Writing}.
\newblock \bibinfo{journal}{\emph{Research in The Teaching of English}}
  \bibinfo{volume}{49} (\bibinfo{year}{2014}), \bibinfo{pages}{95--119}.
\newblock


\bibitem[\protect\citeauthoryear{Paulus}{Paulus}{1999}]%
        {paulus1999}
\bibfield{author}{\bibinfo{person}{Trena~M. Paulus}.}
  \bibinfo{year}{1999}\natexlab{}.
\newblock \showarticletitle{The effect of peer and teacher feedback on student
  writing}.
\newblock \bibinfo{journal}{\emph{Journal of Second Language Writing}}
  \bibinfo{volume}{8}, \bibinfo{number}{3} (\bibinfo{year}{1999}),
  \bibinfo{pages}{265 -- 289}.
\newblock
\showISSN{1060-3743}


\bibitem[\protect\citeauthoryear{Shibani}{Shibani}{2020}]%
        {shibani2020RevisionGraph}
\bibfield{author}{\bibinfo{person}{Antonette Shibani}.}
  \bibinfo{year}{2020}\natexlab{}.
\newblock \showarticletitle{Constructing Automated Revision Graphs: A Novel
  Visualization Technique to Study Student Writing}. In
  \bibinfo{booktitle}{\emph{Artificial Intelligence in Education}}.
  \bibinfo{publisher}{Springer International Publishing},
  \bibinfo{address}{Cham}, \bibinfo{pages}{285--290}.
\newblock
\showISBNx{978-3-030-52240-7}


\bibitem[\protect\citeauthoryear{Shibani, Knight, and Buckingham~Shum}{Shibani
  et~al\mbox{.}}{2018}]%
        {shibani2018kb}
\bibfield{author}{\bibinfo{person}{Antonette Shibani}, \bibinfo{person}{Simon
  Knight}, {and} \bibinfo{person}{Simon Buckingham~Shum}.}
  \bibinfo{year}{2018}\natexlab{}.
\newblock \showarticletitle{Understanding Revisions in Student Writing Through
  Revision Graphs}. In \bibinfo{booktitle}{\emph{International Conference on
  Artificial Intelligence in Education}}. \bibinfo{publisher}{Springer
  International Publishing}, \bibinfo{address}{Cham},
  \bibinfo{pages}{332--336}.
\newblock
\showISBNx{978-3-319-93846-2}


\bibitem[\protect\citeauthoryear{Shute}{Shute}{2008}]%
        {shute2008focus}
\bibfield{author}{\bibinfo{person}{Valerie~J Shute}.}
  \bibinfo{year}{2008}\natexlab{}.
\newblock \showarticletitle{Focus on formative feedback}.
\newblock \bibinfo{journal}{\emph{Review of educational research}}
  \bibinfo{volume}{78}, \bibinfo{number}{1} (\bibinfo{year}{2008}),
  \bibinfo{pages}{153--189}.
\newblock


\bibitem[\protect\citeauthoryear{Southavilay, Yacef, Reimann, and
  Calvo}{Southavilay et~al\mbox{.}}{2013}]%
        {southavilay2013revisionmap}
\bibfield{author}{\bibinfo{person}{Vilaythong Southavilay},
  \bibinfo{person}{Kalina Yacef}, \bibinfo{person}{Peter Reimann}, {and}
  \bibinfo{person}{Rafael~A. Calvo}.} \bibinfo{year}{2013}\natexlab{}.
\newblock \showarticletitle{Analysis of Collaborative Writing Processes Using
  Revision Maps and Probabilistic Topic Models}. In
  \bibinfo{booktitle}{\emph{Proceedings of the Third International Conference
  on Learning Analytics and Knowledge}} (Leuven, Belgium)
  \emph{(\bibinfo{series}{LAK '13})}. \bibinfo{publisher}{Association for
  Computing Machinery}, \bibinfo{address}{New York, NY, USA},
  \bibinfo{pages}{38–47}.
\newblock
\showISBNx{9781450317856}


\bibitem[\protect\citeauthoryear{Toulmin}{Toulmin}{2003}]%
        {toulmin_2003}
\bibfield{author}{\bibinfo{person}{Stephen~E. Toulmin}.}
  \bibinfo{year}{2003}\natexlab{}.
\newblock \bibinfo{booktitle}{\emph{The Uses of Argument} (\bibinfo{edition}{2}
  ed.)}.
\newblock \bibinfo{publisher}{Cambridge University Press}.
\newblock


\bibitem[\protect\citeauthoryear{Turnitin}{Turnitin}{2014}]%
        {turnitin}
\bibfield{author}{\bibinfo{person}{Turnitin}.} \bibinfo{year}{2014}\natexlab{}.
\newblock \showarticletitle{http://turnitin.com/. [Online; accessed
  01-12-2021].}
\newblock  (\bibinfo{year}{2014}).
\newblock
\urldef\tempurl%
\url{http://turnitin.com/}
\showURL{%
\tempurl}


\bibitem[\protect\citeauthoryear{Verbert, Duval, Klerkx, Govaerts, and
  Santos}{Verbert et~al\mbox{.}}{2013}]%
        {verbert2013-learning-analytics}
\bibfield{author}{\bibinfo{person}{Katrien Verbert}, \bibinfo{person}{Erik
  Duval}, \bibinfo{person}{Joris Klerkx}, \bibinfo{person}{Sten Govaerts},
  {and} \bibinfo{person}{José~Luis Santos}.} \bibinfo{year}{2013}\natexlab{}.
\newblock \showarticletitle{Learning Analytics Dashboard Applications}.
\newblock \bibinfo{journal}{\emph{American Behavioral Scientist}}
  \bibinfo{volume}{57}, \bibinfo{number}{10} (\bibinfo{year}{2013}),
  \bibinfo{pages}{1500--1509}.
\newblock


\bibitem[\protect\citeauthoryear{Wambsganss, Niklaus, Cetto, S\"{o}llner,
  Handschuh, and Leimeister}{Wambsganss et~al\mbox{.}}{2020}]%
        {wambsganss2020CHIargskills}
\bibfield{author}{\bibinfo{person}{Thiemo Wambsganss},
  \bibinfo{person}{Christina Niklaus}, \bibinfo{person}{Matthias Cetto},
  \bibinfo{person}{Matthias S\"{o}llner}, \bibinfo{person}{Siegfried
  Handschuh}, {and} \bibinfo{person}{Jan~Marco Leimeister}.}
  \bibinfo{year}{2020}\natexlab{}.
\newblock \showarticletitle{AL: An Adaptive Learning Support System for
  Argumentation Skills}. In \bibinfo{booktitle}{\emph{Proceedings of the 2020
  CHI Conference on Human Factors in Computing Systems}} (Honolulu, HI, USA)
  \emph{(\bibinfo{series}{CHI ’20})}. \bibinfo{publisher}{Association for
  Computing Machinery}, \bibinfo{address}{New York, NY, USA},
  \bibinfo{pages}{1–14}.
\newblock
\showISBNx{9781450367080}


\bibitem[\protect\citeauthoryear{Wang, Olson, Zhang, Nguyen, and Olson}{Wang
  et~al\mbox{.}}{2015}]%
        {wang2015docuviz}
\bibfield{author}{\bibinfo{person}{Dakuo Wang}, \bibinfo{person}{Judith~S.
  Olson}, \bibinfo{person}{Jingwen Zhang}, \bibinfo{person}{Trung Nguyen},
  {and} \bibinfo{person}{Gary~M. Olson}.} \bibinfo{year}{2015}\natexlab{}.
\newblock \showarticletitle{DocuViz: Visualizing Collaborative Writing}. In
  \bibinfo{booktitle}{\emph{Proceedings of the 33rd Annual ACM Conference on
  Human Factors in Computing Systems}} (Seoul, Republic of Korea)
  \emph{(\bibinfo{series}{CHI ’15})}. \bibinfo{publisher}{Association for
  Computing Machinery}, \bibinfo{address}{New York, NY, USA},
  \bibinfo{pages}{1865–1874}.
\newblock
\showISBNx{9781450331456}


\bibitem[\protect\citeauthoryear{Wang, Arya, Novielli, Cheng, and Guo}{Wang
  et~al\mbox{.}}{2020}]%
        {wang2020CHIargulens}
\bibfield{author}{\bibinfo{person}{Wenting Wang}, \bibinfo{person}{Deeksha
  Arya}, \bibinfo{person}{Nicole Novielli}, \bibinfo{person}{Jinghui Cheng},
  {and} \bibinfo{person}{Jin~L.C. Guo}.} \bibinfo{year}{2020}\natexlab{}.
\newblock \showarticletitle{ArguLens: Anatomy of Community Opinions On
  Usability Issues Using Argumentation Models}. In
  \bibinfo{booktitle}{\emph{Proceedings of the 2020 CHI Conference on Human
  Factors in Computing Systems}} (Honolulu, HI, USA)
  \emph{(\bibinfo{series}{CHI ’20})}. \bibinfo{publisher}{Association for
  Computing Machinery}, \bibinfo{address}{New York, NY, USA},
  \bibinfo{pages}{1–14}.
\newblock
\showISBNx{9781450367080}


\bibitem[\protect\citeauthoryear{Wingate}{Wingate}{2010}]%
        {wingate2010impact}
\bibfield{author}{\bibinfo{person}{Ursula Wingate}.}
  \bibinfo{year}{2010}\natexlab{}.
\newblock \showarticletitle{The impact of formative feedback on the development
  of academic writing}.
\newblock \bibinfo{journal}{\emph{Assessment \& Evaluation in Higher
  Education}} \bibinfo{volume}{35}, \bibinfo{number}{5} (\bibinfo{year}{2010}),
  \bibinfo{pages}{519--533}.
\newblock


\bibitem[\protect\citeauthoryear{Writing~Mentor}{Writing~Mentor}{2016}]%
        {ets-writing-mentor}
\bibfield{author}{\bibinfo{person}{The Writing~Mentor}.}
  \bibinfo{year}{2016}\natexlab{}.
\newblock \bibinfo{title}{{ETS} Writing Mentor, https://mentormywriting.org/,
  [Online; accessed 01-12-2021].}
\newblock
\newblock


\bibitem[\protect\citeauthoryear{Yang, Halfaker, Kraut, and Hovy}{Yang
  et~al\mbox{.}}{2017}]%
        {yang2017hkh}
\bibfield{author}{\bibinfo{person}{Diyi Yang}, \bibinfo{person}{Aaron
  Halfaker}, \bibinfo{person}{Robert~E. Kraut}, {and}
  \bibinfo{person}{Eduard~H. Hovy}.} \bibinfo{year}{2017}\natexlab{}.
\newblock \showarticletitle{Identifying Semantic Edit Intentions from Revisions
  in Wikipedia}. In \bibinfo{booktitle}{\emph{Proceedings of the 2017
  Conference on Empirical Methods in Natural Language Processing}}
  \emph{(\bibinfo{series}{EMNLP'17})}. \bibinfo{publisher}{Association for
  Computational Linguistics}, \bibinfo{address}{Copenhagen, Denmark},
  \bibinfo{pages}{9--11}.
\newblock


\bibitem[\protect\citeauthoryear{Zhang, Hashemi, Hwa, and Litman}{Zhang
  et~al\mbox{.}}{2017}]%
        {zhang2017hh}
\bibfield{author}{\bibinfo{person}{Fan Zhang}, \bibinfo{person}{Homa Hashemi},
  \bibinfo{person}{Rebecca Hwa}, {and} \bibinfo{person}{Diane Litman}.}
  \bibinfo{year}{2017}\natexlab{}.
\newblock \showarticletitle{A Corpus of Annotated Revisions for Studying
  Argumentative Writing}. In \bibinfo{booktitle}{\emph{Proceedings of the 55th
  Annual Meeting of the Association for Computational Linguistics (Volume 1:
  Long Papers)}} (Vancouver, Canada). \bibinfo{publisher}{Association for
  Computational Linguistics}, \bibinfo{pages}{1568--1578}.
\newblock


\bibitem[\protect\citeauthoryear{Zhang, Hwa, Litman, and B.~Hashemi}{Zhang
  et~al\mbox{.}}{2016}]%
        {zhang2016hl}
\bibfield{author}{\bibinfo{person}{Fan Zhang}, \bibinfo{person}{Rebecca Hwa},
  \bibinfo{person}{Diane Litman}, {and} \bibinfo{person}{Homa B.~Hashemi}.}
  \bibinfo{year}{2016}\natexlab{}.
\newblock \showarticletitle{ArgRewrite: A Web-based Revision Assistant for
  Argumentative Writings}. In \bibinfo{booktitle}{\emph{Proceedings of the 2016
  Conference of the North American Chapter of the Association for Computational
  Linguistics: Demonstrations}}. \bibinfo{publisher}{Association for
  Computational Linguistics}, \bibinfo{address}{San Diego, California},
  \bibinfo{pages}{37--41}.
\newblock


\bibitem[\protect\citeauthoryear{Zhang and Litman}{Zhang and Litman}{2014}]%
        {zhang2014alignment}
\bibfield{author}{\bibinfo{person}{Fan Zhang} {and} \bibinfo{person}{Diane
  Litman}.} \bibinfo{year}{2014}\natexlab{}.
\newblock \showarticletitle{Sentence-level Rewriting Detection}. In
  \bibinfo{booktitle}{\emph{Proceedings of the Ninth Workshop on Innovative Use
  of {NLP} for Building Educational Applications}}.
  \bibinfo{publisher}{Association for Computational Linguistics},
  \bibinfo{address}{Baltimore, Maryland}, \bibinfo{pages}{149--154}.
\newblock


\bibitem[\protect\citeauthoryear{Zhang and Litman}{Zhang and Litman}{2015}]%
        {zhang2015l}
\bibfield{author}{\bibinfo{person}{Fan Zhang} {and} \bibinfo{person}{Diane
  Litman}.} \bibinfo{year}{2015}\natexlab{}.
\newblock \showarticletitle{Annotation and Classification of Argumentative
  Writing Revisions}. In \bibinfo{booktitle}{\emph{Proceedings of the 10th
  Workshop on Innovative Use of NLP for Building Educational Applications}}.
  \bibinfo{publisher}{Association for Computational Linguistics},
  \bibinfo{address}{Denver, Colorado}, \bibinfo{pages}{133--143}.
\newblock


\bibitem[\protect\citeauthoryear{Zimmerman and Kitsantas}{Zimmerman and
  Kitsantas}{2002}]%
        {ZimmermanKitsantas2002}
\bibfield{author}{\bibinfo{person}{Barry Zimmerman} {and}
  \bibinfo{person}{Anastasia Kitsantas}.} \bibinfo{year}{2002}\natexlab{}.
\newblock \showarticletitle{Acquiring writing revision and self-regulatory
  skill through observation and emulation}.
\newblock \bibinfo{journal}{\emph{Journal of Educational Psychology}}
  \bibinfo{volume}{94} (\bibinfo{date}{12} \bibinfo{year}{2002}),
  \bibinfo{pages}{660--668}.
\newblock


\bibitem[\protect\citeauthoryear{Zimmerman and Bandura}{Zimmerman and
  Bandura}{1994}]%
        {ZimmermanBandura1994}
\bibfield{author}{\bibinfo{person}{Barry~J. Zimmerman} {and}
  \bibinfo{person}{Albert Bandura}.} \bibinfo{year}{1994}\natexlab{}.
\newblock \showarticletitle{Impact of Self-Regulatory Influences on Writing
  Course Attainment}.
\newblock \bibinfo{journal}{\emph{American Educational Research Journal}}
  \bibinfo{volume}{31}, \bibinfo{number}{4} (\bibinfo{year}{1994}),
  \bibinfo{pages}{845--862}.
\newblock


\end{thebibliography}

\appendix

\section{Data Collection Materials}
\label{appendix}

\subsection{Prompt}
\label{a sec: prompt and rubric}


In this argumentative writing task, imagine that you are writing an op-ed piece for the Pittsburgh City Paper about self-driving cars. The editor of the paper has asked potential writers, like you, to gather information about the use of self-driving cars, and argue whether they are beneficial or not beneficial to society. In your writing, first, briefly explain both the advantages and disadvantages of self-driving cars. Then, you will choose a side, and construct an argument in support of self-driving cars as beneficial to society, or against self-driving cars as not beneficial to society.

A high quality op-ed piece maintains a clear position on the issue and uses supporting ideas, strong evidence from the reading, explanations of your ideas and evidence, and a counter-argument. Furthermore, a high quality op-ed piece is clearly organized, uses precise word choices, and is grammatically correct.

\subsection{Example of Expert Feedback on Draft1}
\label{a sec: expert feedback}

Thank you for your participation in the study. Your draft has been read, and feedback from an expert writing instructor is written below. We advise that you use this feedback when you revise. 

The strengths of your essay include:
\begin{itemize}
    \item All claims have relevant supporting evidence, though that evidence may be brief or general.
    \item You respond to one, but not all parts of the prompt. However, your entire essay is focused on the prompt.
\end{itemize}
Areas to improve in your essay include:  
\begin{itemize}
    \item You provided a statement that somewhat show your stance for or against self-driving cars, but it is unclear, or is just a restatement of the prompt.
    \item Your essay’s sequence of ideas is inconsistent, with some clear and some unclear progression. 
    \item Your essay does not include a rebuttal.
\end{itemize}

\subsection{Scoring Rubric}
\label{a sec: rubric}
Table~\ref{a tab:rubric} shows the scoring rubric used to provide feedback.

\clearpage
\onecolumn
\begin{longtable}{p{0.1\linewidth}|p{0.15\linewidth}|p{0.15\linewidth}|p{0.18\linewidth}|p{0.18\linewidth}}
    \caption[Argumentative Essay Scoring Rubric]{Argumentative Essay Rubric}
    \label{a tab:rubric} \\
    &   \textbf{1-Poor} & \textbf{2-Developing} & \textbf{3-Proficient} & \textbf{4-Excellent} \\ 
     \hline
     \endfirsthead
    &   \textbf{1-Poor} & \textbf{2-Developing} & \textbf{3-Proficient} & \textbf{4-Excellent} \\ 
    \hline
     \endhead
      \textbf{Response to prompt}
     & 
         The essay is off topic, and does not consider or respond to the prompt in any way. &
         The essay addresses the topic, but the entire essay is not focused on the prompt. The author may get off topic at points. &
         The author responds to one, but not all parts of the prompt, but the entire essay is focused on the prompt. &
         The author responds to all parts of the prompt and the entire essay is focused on the prompt. \\ \hline
         
     \textbf{Thesis} &
         The author did not include a statement that clearly showed the author’s stance for or against selfdriving cars. &
         The author provided a statement that somewhat showed the author’s stance for or against self-driving cars, though it may be unclear or only a restatement of the essay prompt. &
         The author provided a brief statement that reflects a thesis, and is indicative of the stance the author is taking toward self-driving cars. &
         The author provided a clear, nuanced and original statement that acted as a specific stance for or against self-driving cars. \\  \hline
         
     \textbf{Claims} & 
        The author’s claims are difficult to understand or locate. & 
        The author’s claims are present, but are unclear, not fully connected to the thesis or the reading, or the author makes only one claim multiple times. & 
        The author makes multiple, distinct, and clear claims that align with either their thesis or the given reading, but not both. & 
        The author makes multiple, distinct claims that are clear, and align with both their thesis statement and the given reading. They fully support the author’s argument. \\  \hline
        
     \textbf{Evidence for Claims} & 
        The author does not provide any evidence to support thesis/claims. & 
        Less than half of claims are supported with relevant or credible evidence or the connections between the evidence and the thesis/claims is not clear. & 
        All claims have relevant supporting evidence, though that evidence may be brief or general. The source of the evidence is credible and acknowledged/cited where appropriate. & 
        The author provides specific and convincing evidence for each claim, and most evidence is given through detailed personal examples, relevant direct quotations, or detailed examples from the provided reading. The source of the evidence is credible and acknowledged/cited where appropriate. \\  \hline
        
     \textbf{Reasoning} & 
        The author provides no reasoning for any of their claims. & 
        Less than half of claims are supported with reasoning or the reasoning is so brief, it essentially repeats the claim. Some reasoning may not appear logical or clear. & 
        All claims are supported with reasoning that connect the evidence to the claim, though some may not be fully explained or difficult to follow. & 
        All claims are supported with clear reasoning that shows thoughtful, elaborated analysis. \\  \hline
    
    \textbf{Reordering/ Organization} &
    	The sequence of ideas/claims is difficult to follow and the essay does not have an introduction, conclusion, and body paragraphs that are organized clearly around distinct claims. &
    	The essay’s sequence of ideas is inconsistent, with some clear and some unclear progression of ideas OR the essay is missing a distinct introduction OR conclusion. &
    	The essay has a clear introduction, body, and conclusion and a logical sequence of ideas, but each claim is not located in its own separate paragraph. &
    	The essay has an introduction, body and conclusion and a logical sequence of ideas. Each paragraph makes a distinct claim. \\ \hline
      
    \textbf{Rebuttal} &
    	The essay does not include a rebuttal. &
    	The essay includes a rebuttal in the sense that it acknowledges another point of view, but does not explore possible reasons why this other viewpoint exists. &
    	The essay includes a rebuttal in the form of an acknowledgement of a different point of view and reasons for that view, but does not explain why those reasons are incorrect or unconvincing. &
    	The essay explains a different point of view and elaborates why it is not convincing or correct. \\ \hline
    	
    \textbf{Precision} &
    	Throughout the essay, word choices are overly informal and general (e.g., “I don’t like self-driving cars because they have problems.”). &
    	Word choices are mostly overly general and informal, though at times they are specific. &
    	Word choices are mostly specific though there may be a few word choices that make the meaning of the sentence vague. &
    	Throughout the essay, word choices are specific and convey precise meanings (e.g., “Self-driving cars are dangerous because the technology is still not advanced enough to address the ethical decisions drivers must make.”) \\ \hline
    
    \textbf{Fluency} &
    	A majority of sentences are difficult to understand because of incorrect/ inappropriate word choices and sentence structure. &
    	A noticeable number of sentences are difficult to understand because of incorrect/ inappropriate word choices and sentence structure, although the author’s overall point is understandable. & 
    	Most sentences are clear because of correct and appropriate word choices and sentence structure. &
    	All sentences are clear because of correct and appropriate word choices and sentence structure. \\ \hline
    	
    \textbf{Conventions/ Grammar/ Spelling} &
    	The author makes many grammatical or spelling errors throughout their piece that interfere with the meaning. &
    	The author makes many grammatical or spelling errors throughout their piece, though the errors rarely interfere with meaning. &
    	The author makes few grammatical or spelling errors throughout their piece, and the errors do not interfere with meaning. &
    	The author makes few or no grammatical or spelling errors throughout their piece, and the meaning is clear. \\
    \hline   
\end{longtable}
\clearpage
\twocolumn
\end{document}